\documentclass[twocolumn,twocolappendix]{aastex701}
\usepackage{physics}
\usepackage{enumitem}
\usepackage{CJK}

\graphicspath{{./}{}}

\newcommand{\F}{\mathbf{F}_{r}}
\newcommand{\G}{\mathbb{G}}

\newcommand{\kr}{\kappa_{\rm{R}}}
\newcommand{\eg}{e_{\rm g}}
\newcommand{\heating}{\dot{\Lambda}}
\newcommand{\vel}{\mathbf{v}}

\newcommand{\mpl}{m_\text{p}}
\newcommand{\rpl}{r_\text{p}}
\newcommand{\macc}{\dot{m}_{\text{acc}}}
\newcommand{\mflux}{\dot{m}_{\text{flux}}}
\newcommand{\rout}{r_{\text{out}}}
\newcommand{\rcirc}{\bar{r}_{\text{circ}}}
\newcommand{\tvis}{t_{\text{vis}}}
\newcommand{\rj}{R_{\rm J}}
\newcommand{\mj}{M_{\rm J}}
\newcommand{\rin}{r_{\rm{in}}}
\newcommand{\cmsg}{cm$^{2}\cdot$g$^{-1}$}
\newcommand{\gcmc}{g\,cm$^{-3}$}
\newcommand{\mbacc}{\dot{\bar{m}}_{\rm acc}}

\definecolor{THUPurple}{RGB}{101,0,153}

\shorttitle{Thermodynamics of Viscously Heated CPD}
\shortauthors{Chen, Yang \& Bai}

\begin{document}
\begin{CJK*}{UTF8}{gbsn}
\title{Thermodynamics of Viscously Heated Circumplanetary Disks}

\author[orcid=0000-0001-7420-9606]{Zhuo Chen (陈卓)}
\affiliation{Chen-Ning Yang Institute for Advanced Study, Tsinghua University, Beijing 100084, P. R. China}
\email[show]{chenzhuo\_astro@tsinghua.edu.cn}

\author[0009-0008-1149-4217]{Jing Yang (杨晶)}
\affiliation{Chen-Ning Yang Institute for Advanced Study, Tsinghua University, Beijing 100084, P. R. China}
\email{jing-yan24@mails.tsinghua.edu.cn}

\author[0000-0001-6906-9549]{Xue-Ning Bai (白雪宁)}
\affiliation{Chen-Ning Yang Institute for Advanced Study, Tsinghua University, Beijing 100084, P. R. China}
\affiliation{Department of Astronomy, Tsinghua University, Beijing 100084, P. R. China}
\email[show]{xbai@mail.tsinghua.edu.cn}

\begin{abstract}
Circumplanetary disks (CPDs) play an important role in the formation of gas giants and their satellites, yet remain poorly understood owing to the lack of observables and comprehensive theoretical models. Incorporating $\alpha$-viscosity, a free-fall prescription of mass infall and boundary layer accretion, we develop an axisymmetric viscous radiation hydrodynamic (RHD) model. We evolve this model for different physical parameters using the RHD code {\tt Guangqi}, which solves RHD equations under the flux-limited diffusion approximation. In eight out of nine simulations, we obtain quasi-steady-state solutions and verify self-consistent energy conversion and transport. Overall, disk thickness increases with accretion rate and decreases with planet mass, consistent with the expected trend in conventional $\alpha$-disk models. On the other hand, we highlight that the boundary layer plays a significant role in setting angular momentum transport and thermodynamics of the inner CPD, yielding ``hot'' ($s_{\rm real}>11\,k_{\rm{B}}\,m_{\rm{H}}^{-1}$) accreted gas, and potentially leading to dust sublimation. We identify a supersonic accretion flow, amounting to $2\%$--$5\%$ of the total accretion rate, located at high elevations above the CPDs with low angular momentum in our simulations. If the planet is massive ($\gtrsim3\,M_{\rm J}$), this low angular momentum flow impacting the planetary surface could generate observable H$\alpha$ emission when the shock accretion rate is $\gtrsim5.5\times10^{-8}\, M_{\rm J}\,\rm{yr}^{-1}$.
\end{abstract}


\section{Introduction}

High-resolution imaging from facilities such as ALMA has transformed our understanding of protoplanetary disks (PPDs), the birthplaces of planets \citep{andrews2020}. These observations routinely reveal complex substructures---rings, gaps, and cavities \citep{andrews2018,huang2018,long2018}---widely interpreted as the first observable signposts of ongoing planet formation \citep{zhang2018}. Such features reflect the dynamical imprint of a massive protoplanet sculpting its natal environment: as the planet grows, angular momentum exchange with the disk repels material along its orbit and opens a gap \citep{lin1986,crida2006,fung2016}. Although this isolates the planet from its local reservoir, material from the wider PPD streams across the gap and sustains its growth. This inflowing gas and dust, carrying substantial angular momentum, is expected to settle into a compact, rotationally supported accretion disk: the circumplanetary disk (CPD). The CPD regulates the final accretion of mass onto the planet and is the primary site for satellite formation, particularly the Galilean satellites around Jupiter \citep{canup2002}.

Among the candidate protoplanets identified within PPDs \citep{keppler2018,hammond2023,close2025a,currie2025,vancapelleveen2025}, PDS 70 remains the unambiguous system that has forming planets hosting CPDs. The detection of H$\alpha$ emission from PDS 70b indicates active accretion driven by circumplanetary material \citep{wagner2018,haffert2019,hashimoto2020,zhou2021,close2025b}. The presence of a CPD around PDS 70c is supported by multi-wavelength observations from ALMA \citep{benisty2021,doi2024,fasano2025,dominguez2025}, VLT \citep{wang2021}, and JWST \citep{blakely2025}. Using comprehensive spectral energy distribution (SED) modeling, \cite{wang2021} placed an upper limit of 0.3\,AU on the size of the CPD of PDS 70b, corresponding to approximately 0.2 Hill radii ($R_{\rm H}$). Despite these advances, the physical properties of CPDs remain poorly constrained; fundamental quantities such as surface density, temperature structure, and rotation profiles are currently beyond observational reach.

To investigate the formation environment of these gas giants, numerous three-dimensional (3D) multi-physics simulations of accreting planets have been performed. \cite{tanigawa2012} first showed that gas may stream through the gap from the L1 and L2 points toward the planet, entering from high latitudes with non-zero angular momentum, and subsequent high-resolution 3D simulations have confirmed this flow pattern \citep{fung2016,maeda2022,szulagyi2022,li2023,lega2024,sagnbayeva2025}. While non-ideal magnetohydrodynamic (MHD) simulations suggest that magnetic torques help open wider gaps \citep{aoyama2023,wafflard2023,hu2025}, the flow topology in the planetary vicinity remains comparable to the pure-hydrodynamic case when the ionization fraction is low \citep{aoyama2026}. Thermodynamics, in turn, strongly shapes both the flow pattern and the morphology of the circumplanetary material \citep{krapp2024,sagnbayeva2025}.

3D simulations are computationally expensive, which limits their ability to resolve the innermost region of a CPD ($\lesssim0.1\,R_{\rm H}$) where significant hydrodynamic and thermodynamic processes occur.
There, the stellar tidal potential diminishes rapidly toward the planet, so the inner CPD is expected to approach an axisymmetric state---enabling axisymmetric simulations to resolve the disk interior at far lower computational cost.
\cite{takasao2021} and \cite{marleau2023} adopted this geometry to study flow structures and potential shock signatures near the planet, while \cite{dong2021} performed high-resolution axisymmetric viscous hydrodynamic simulations to investigate boundary layer (BL) effects, finding that planetary spin significantly influences the accretion rate. \citet{tang2022,tang2026} investigated the role of magnetic fields in CPD accretion.

\begin{figure}
    \centering
    \includegraphics[width=\linewidth]{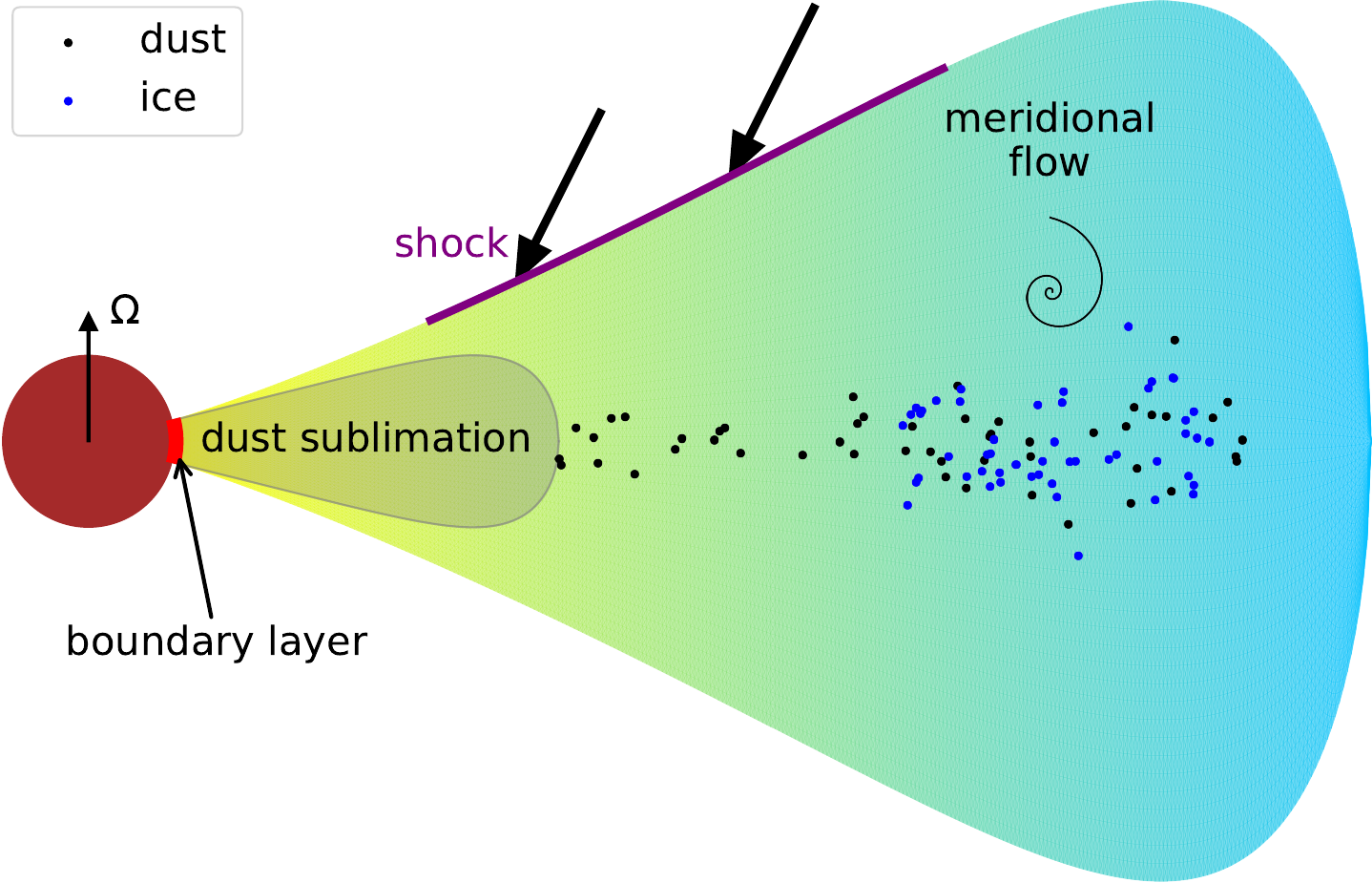}
    \caption{Schematic illustration of the non-magnetized CPD model. Gas accreting from the PPD impacts the CPD surface at high latitude, forming a shock front. The disk exhibits a meridional flow pattern and undergoes viscously-driven accretion.
    A boundary layer at the interface between the accreting planet and the inner disk efficiently converts kinetic energy into heat. This energy release, combined with effective viscous dissipation,
    leads to a radial temperature profile that encompasses not only the water snowline but also sustains a dust sublimation zone in the inner CPD.
    }
    \label{fig:cpd}
\end{figure}

Thermodynamics is a crucial factor controlling the formation and morphology of CPDs \citep{szulagyi2016,krapp2024}, and requires a realistic treatment of radiation transport. In particular, simulations with a locally-isothermal equation of state (EoS) preferentially form CPDs, whereas under an adiabatic EoS the circumplanetary region becomes a quasi-spherical envelope \citep{fung2019}. Thermodynamics further governs the entropy of the accreted gas \citep{owen2016,chen2022}, the chemical composition of the disk \citep{oberg2023,cridland2025}, and satellite formation and migration \citep{canup2006,bierson2020,fujii2020}.
Accurate modeling of CPD thermodynamics requires a self-consistent treatment of accretion heating, compressional work, intense radiative feedback from the boundary layer, and the internal luminosity of the planet; coupling these processes to radiation transport demands high-precision numerical tools capable of quantifying energy conversion and transport throughout the disk.

However, current CPD models struggle to adequately resolve these complex processes. Due to computational limits, 3D RHD models generally cannot resolve the highly thermodynamically active inner region \citep{szulagyi2016,krapp2024}. Conversely, 1D models often omit the BL entirely and rely on simplified radiative cooling prescriptions that do not necessarily yield accurate temperature profiles \citep{taylor2025}.

To address these challenges, we develop an axisymmetric viscous radiation hydrodynamic (RHD) CPD model, as illustrated in Figure \ref{fig:cpd}. The remainder of this paper is organized as follows. Section \ref{sec:model} describes the physical model, derives the governing equations, and outlines the numerical scheme. We then detail the numerical setup, including initial and boundary conditions, in Section \ref{sec:setups}. The simulation results are presented in Section \ref{sec:results}, followed by a discussion of relevant physics and model caveats in Section \ref{sec:discussion}. Finally, we summarize our conclusions in Section \ref{sec:conclusion}.

\section{Physical model}\label{sec:model}

We model the CPD as an axisymmetric accretion flow governed by the system of viscous radiation hydrodynamics (RHD). The simulations are performed in a spherical, irrotational reference frame $(r,\theta)$ that co-moves with the planet. Regarding the physical origin of viscosity, \citet{fujii2014,keith2014} found that the ionization fraction in a dense circumplanetary disk (CPD) is often too low to effectively drive the magneto-rotational instability \citep{balbus1991} or a disk wind \citep{bai2013}. Specifically, the Elsasser number reaches unity above two disk scale heights only if the surface density is below $0.01\rm{~g~cm}^{-2}$ \citep{fujii2014}. On the other hand, recent two-dimensional (2D) and 3D hydrodynamical simulations show that spiral shocks and density waves may give rise to an effective viscosity that amounts to $\alpha\sim10^{-3}$--$10^{-2}$ \citep{zhu2016,wang2026}, which motivates us to adopt effective values of $\alpha=\{10^{-3},10^{-2}\}$ in this work (see Section \ref{sec:discussion} for a discussion of the viscosity in BLs).

A physical model of the planet and planetary atmosphere is also critical for boundary-layer accretion. We model the planetary atmosphere as a rotating sphere by specifying a fixed pressure across the guard cells of the inner boundary. A thin planetary atmosphere emerges due to the balance between the gas pressure gradient and gravity (see Section \ref{sec:ibc} for more details).

\subsection{The Governing Equations}

We start by listing the radiation hydrodynamic equations as follows:
\begin{eqnarray}
    \pdv{\rho}{t}+\nabla\cdot(\rho\vel)&=&0,   \label{eqn:mass}\\
    \pdv{\rho\vel}{t}+\nabla\cdot(\rho\vel\vel-\boldsymbol{\sigma}+p\mathbb{I})&=&\rho\mathbf{g},   \label{eqn:mom}\\
    \pdv{E}{t}+\nabla\cdot[\vel(E+p)-{\boldsymbol{\sigma}\cdot\vel}]&=&\rho\mathbf{g}\cdot\vel+G^{0},   \label{eqn:hydroenergy}\\
    \pdv{E_{r}}{t}+\nabla\cdot\F&=&-G^{0},    \label{eqn:radenergy}    \\
    E&=&\eg+\frac{\rho|\vel|^2}{2},
\end{eqnarray}
where $\rho,\vel,p,\eg$, and $E$ are the density, velocity, pressure, internal energy density, and total energy of the gas, respectively. For ideal gas, $\eg=p/(\gamma-1)$, where $\gamma=1.4$ is the assumed adiabatic index, and $p=nk_{\rm B}T$, where $n=\rho/(\mu m_{\rm H})$ is the particle number density, $\mu=2.3$ is the assumed mean molecular weight, and $m_{\rm H}$ is the mass of a hydrogen atom.

We consider point source gravity {\it without} gravitational softening, with
\begin{equation}
    \mathbf{g}=G\mpl\mathbf{r}/r^{3}
\end{equation}
being the gravitational acceleration. Here, $G,\mpl,\mathbf{r}$ are the gravitational constant, mass of the planet, and radial vector to the planet, respectively.

In an isotropic effective viscous fluid model, the viscous stress tensor is defined as:
\begin{equation}
\boldsymbol{\sigma} = \rho\nu [\nabla \mathbf{v} + (\nabla \mathbf{v})^{T}],
\end{equation}
where $\nu$ is the kinematic viscosity, and we adopt the standard $\alpha$-prescription \citep{shakura1973}
\begin{equation}
\nu = \alpha \frac{c_T^2}{\Omega}=\alpha \frac{p}{\rho \Omega},
\end{equation}
where $c_T=\sqrt{p/\rho}$ is the isothermal sound speed, and $\Omega$ is the local angular frequency. In this CPD model, the $r\phi$ and $\theta\phi$ components of the stress tensor, $\sigma_{r\phi}$ and $\sigma_{\theta\phi}$, are the dominant driver of both thermodynamics and angular momentum transport. For simplicity, we focus exclusively on these two components, and neglect stress components in other directions (see Section \ref{sec:psama} for its application).

To solve the radiation transport equation, we adopt the flux-limited diffusion (FLD) approximation \citep{levermore1981}, which relates the radiation energy density $E_r$ and radiation flux $\mathbf{F}_r$ by:
\begin{align}
\mathbf{F}_r &= -\frac{c\lambda(\mathcal{R})}{\kr\rho} \nabla E_r, \\
\mathcal{R} &= \frac{|\nabla E_r|}{\kappa_{\mathrm{R}}\rho E_r} \label{eqn:R},
\end{align}
where $\kappa_{\mathrm{R}}$ is the Rosseland mean opacity. The flux limiter is given by $\lambda(\mathcal{R})=(2+\mathcal{R})/(6+3\mathcal{R}+\mathcal{R}^2)$ \citep{levermore1981}. The gas and radiation are coupled via the energy source term $G^0$:
\begin{equation}
G^{0} = \kappa_{\mathrm{P}}\rho c (E_r - a_{\mathrm{rad}}T^{4}),
\end{equation}
where $\kappa_{\mathrm{P}}$ is the Planck mean opacity, $a_{\mathrm{rad}}$ is the radiation constant, and $c$ is the speed of light.

We adopt the dust \citep{semenov2003} and gas \citep{malygin2014} opacity model and take the maximum of the interpolated opacity values from the two tables. Considering that dust may grow in the CPD and that the PPD is probably in a metal-poor state after gas giant formation, we tentatively reduce the dust opacity table by a factor of 2.

\subsection{Passive Scalar Angular Momentum Algorithm with Viscous Torques}\label{sec:psama}

We employ the passive scalar angular momentum algorithm (PSAMA) \citep{chen2026}, which replaces the linear momentum equation in the azimuthal direction with the angular momentum equation in a hydrodynamic Riemann solver, to reformulate Equations \ref{eqn:mass}--\ref{eqn:radenergy} in axisymmetric spherical coordinates, incorporating viscous torques. With the simplification that $\sigma_{rr}=\sigma_{r\theta}=\sigma_{\phi r}=\sigma_{\phi\theta}=\sigma_{\theta\theta}=0$, we can express the work done by viscous stress as:
\begin{equation}
\boldsymbol{\sigma}\cdot\mathbf{v} = \Omega\boldsymbol{\tau},
\end{equation}
where $\boldsymbol{\tau}=[\sigma_{r\phi}r\sin\theta,\sigma_{\theta\phi}r\sin\theta,0]^{T}$ represents the viscous torque vector along the $z$-axis.

We reformulate the viscous RHD equations using PSAMA by transforming the momentum conservation equation in the $\phi$ direction into an angular momentum conservation equation. The resulting system in spherical coordinates is:
\begin{eqnarray}
    \pdv{\rho}{t}+\frac{1}{r^{2}}\pdv{(r^{2}\rho v_{r})}{r}+\frac{1}{r\sin\theta}\pdv{(\sin\theta\rho v_{\theta})}{\theta}=0,&&\quad\quad\label{eqn:spheremass}\\
    \pdv{(\rho v_{r})}{t}+\frac{1}{r^2}\pdv{[r^{2}(\rho v^{2}_{r}+p)]}{r}+\frac{1}{r\sin\theta}\pdv{(\sin\theta\rho v_{r}v_{\theta})}{\theta}=&&\notag\\
    \rho g_{r}+\frac{2p}{r}+\frac{\rho(v^{2}_{\theta}+v^{2}_{\phi})}{r},&&\label{eqn:spherevr}\\
    \pdv{(\rho v_{\theta})}{t}+\frac{1}{r^2}\pdv{(r^{2}\rho v_{r}v_{\theta})}{r}+\frac{1}{r\sin\theta}\pdv{[\sin\theta(\rho v^{2}_{\theta}+p)]}{\theta}=&&\notag\\
    \frac{\cot\theta(\rho v^{2}_{\phi}+p)-\rho  v_{r}v_{\theta}}{r},&&\label{eqn:spherevtheta}\\
    \pdv{E}{t}+\nabla\cdot{[(E+p)\vel]}=\rho g_{r}v_{r}+\nabla\cdot(\Omega\boldsymbol{\tau})+G^{0},&&\label{eqn:totalE}\\
    \pdv{(\rho l)}{t}+\nabla\cdot(\rho l\vel)=\nabla\cdot\boldsymbol{\tau},\label{eqn:spheream}
\end{eqnarray}
where $\vel=[v_{r}, v_{\theta}, v_{\phi}]^T$ is the velocity in the spherical coordinate and $l=\Omega(r\sin\theta)^{2}$ is the specific angular momentum.

We then decompose the total energy density $E$ into a poloidal component $E'$ and the azimuthal kinetic energy $e_{k\phi}$:
\begin{equation}
    E=\underbrace{\eg+\rho\bigg(\frac{v_{r}^{2}+v_{\theta}^{2}}{2}\bigg)}_{E'}+\underbrace{\frac{\rho l^{2}}{2(r\sin\theta)^{2}}}_{e_{k\phi}}\ .
\end{equation}
Following the derivation in \cite{chen2026}, the evolution of the rotational kinetic energy in the presence of viscous torque is given by:
\begin{equation}\label{eqn:ekphi}
    \pdv{e_{k\phi}}{t}+\div{(e_{k\phi}\vel)}=-\rho\vec{f}_{\Omega}\cdot\vel+\Omega\div{\boldsymbol{\tau}},
\end{equation}
where $\vec{f}_{\Omega}=r\Omega^{2}\sin\theta[\sin\theta,\cos\theta,0]^{T}$ is the centrifugal force. Subtracting Equation \ref{eqn:ekphi} from Equation \ref{eqn:totalE} yields the evolution equation for $E'$ subject to the viscous torque $\boldsymbol{\tau}$:
\begin{equation}\label{eqn:E'}
    \pdv{E'}{t}+\nabla\cdot{[(E'+p)\vel]}=\rho g_{r}v_{r}+\boldsymbol{\tau}\cdot\nabla\Omega+G^{0}+\rho\vec{f}_{\Omega}\cdot\vel.
\end{equation}
Specifically, the term $\boldsymbol{\tau}\cdot\nabla\Omega$ represents the effective viscous heating rate \citep{landau1987}:
\begin{align}\label{eqn:heating}
    \boldsymbol{\tau}\cdot\nabla\Omega=&r\sin\theta\bigg(\sigma_{r\phi}\pdv{\Omega}{r}+\sigma_{\theta\phi}\pdv{\Omega}{\theta}\bigg)    \notag\\
    =&\frac{1}{2}\boldsymbol{\sigma}:[\nabla\vel+(\nabla\vel)^{T}]=\heating.
\end{align}
The viscous heating source term $\heating$ converts kinetic energy into internal energy $\eg$. In the BL, $\heating$ can become an energy-stiff source term and must be solved implicitly alongside the radiation transport equation. Therefore, we solve Equation \ref{eqn:E'} for $E'$ in two steps. The non-stiff source terms are integrated in the explicit hydrodynamic solver:
\begin{equation}\label{eqn:sphereE}
    \pdv{E'}{t}+\nabla\cdot{[(E'+p)\vel]}=\rho g_{r}v_{r}+\rho\vec{f}_{\Omega}\cdot\vel,
\end{equation}
while the stiff terms are integrated implicitly together with the radiation transport:
\begin{align}
    \pdv{\eg}{t}=&G^{0}+\heating,\label{eqn:radeg}\\
    \pdv{E_{r}}{t}+\nabla\cdot\F=&-G^{0}.\label{eqn:rader}
\end{align}

In summary, the 2D viscous RHD disk model is solved in four main steps using the code {\tt Guangqi} \citep{chen2026}:
\begin{enumerate}[leftmargin=15pt]
\item Solve the hydrodynamic Equations \ref{eqn:spheremass}, \ref{eqn:spherevr}, \ref{eqn:spherevtheta}, and \ref{eqn:sphereE} explicitly.
\item Compute the viscous torque $\boldsymbol{\tau}$ and heating rate $\heating$.
\item Evolve the angular momentum Equation \ref{eqn:spheream} as a passive scalar to strictly conserve angular momentum.
\item Solve the radiation transport Equations \ref{eqn:radeg} and \ref{eqn:rader} using a fully implicit solver.
\end{enumerate}
These steps constitute the implementation of the PSAMA in the presence of viscous torques.

\section{Simulation setups}\label{sec:setups}

\subsection{Computational Domain Setups}\label{sec:computationaldomain}

\begin{figure}
    \centering
    \includegraphics[width=\columnwidth]{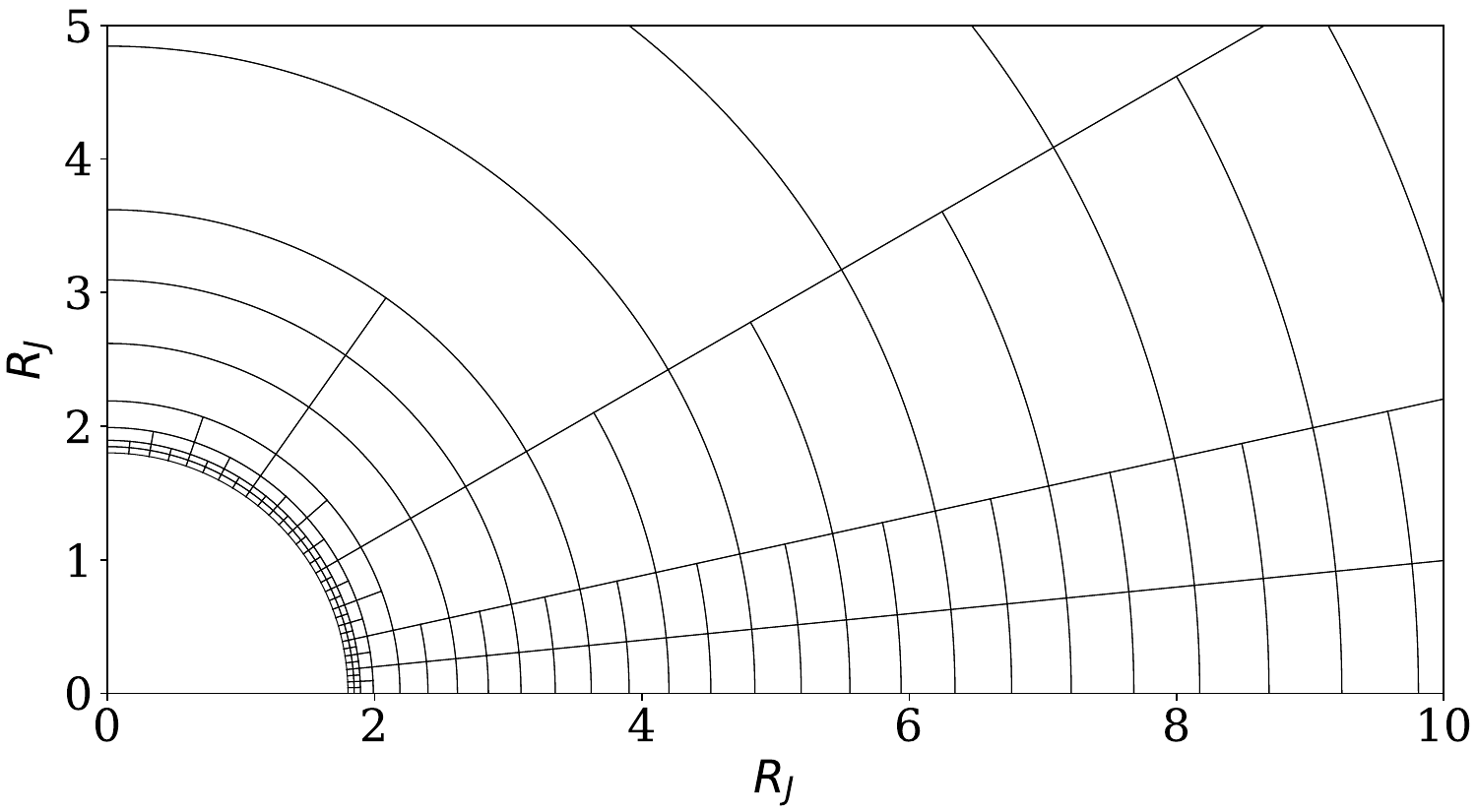}
    \caption{The block structure of the stretched grid with mesh refinement of model 7 in Table \ref{tab:bc}. Each block consists of $32\times32$ cells.}
    \label{fig:cd}
\end{figure}

The simulations are performed using {\tt Guangqi}, a 2D Godunov code with block-based AMR available for both Cartesian and spherical polar coordinates \citep{chen2026}. Our computational domain spans $[\rpl, 50\rpl] \times [0, \pi/2]$ in the radial and polar directions, respectively, where $\rpl$ is the assumed planetary radius. We impose mirror symmetry at the equator.

The base resolution is $256 \times 32$, organized into blocks of $32 \times 32$ cells. To capture critical scales, we apply up to 5 levels of static mesh refinement (SMR) to resolve the planetary atmosphere near the inner boundary, and 3 levels of SMR to resolve the CPD near the equator. {\tt Guangqi} doubles the resolution in all coordinates for each level of refinement.

To enhance computational efficiency, we adopt a geometrically stretched grid along both the $r$ and $\theta$ coordinates. Figure \ref{fig:cd} illustrates an example of the stretched grid structure with mesh refinement. The grid spacing is defined by:
\begin{align}
\Delta r_{i+1} &= q_{r}\Delta r_{i}, \\
\Delta\theta_{j+1} &= q_{\theta}\Delta\theta_{j},
\end{align}
where $i$ and $j$ are the indices along the $r$ and $\theta$ coordinates, with stretching factors $q_{r}=1.01265$ and $q_{\theta}=0.95760$.

This configuration yields a minimum cell size of $7.8 \times 10^{-4} \rpl$ in the radial direction.
This is to be compared with the scale height of the (isothermal) planetary atmosphere $H_a$, which can be found to be
\begin{equation}
H_{a} = 0.01 \cdot \left(\frac{\rpl}{1.98\,\rj}\right)^2 \cdot \left(\frac{\mj}{\mpl}\right) \cdot \left(\frac{T}{1400\,\rm{K}}\right) \rj.
\end{equation}
For the case where $\rpl=1.8\,\rj$, our resolution provides approximately 5--6 cells per $H_{a}$. The smallest angular cell size near the mid-plane is $4.85 \times 10^{-3}$ rad, which is 15 times smaller than the minimum CPD aspect ratio ($H/r > 0.075$; see Section \ref{sec:cpecpd}), and we consider this resolution sufficient to resolve our models.

We set the CFL number to 0.6 for all simulations. Additionally, we adopt 2 sub-timesteps to improve the temporal resolution of the radiation transport solver ($n_{\text{iter}}=2$; see \citet{chen2026} for details).

\subsection{Initial and Boundary Conditions}\label{sec:ibc}

\begin{table*}[]
    \centering
    \begin{tabular}{ccccc|cccc|cc|cc}\hline\hline
        \multicolumn{5}{c|}{models} & \multicolumn{4}{c|}{outer boundary}   & \multicolumn{2}{c|}{inner} & \multicolumn{2}{c}{results} \\\hline
        No. & description & $\mpl$ & $\rpl$ & $\alpha$ & $\theta_{0}$ & $\rho_{0}$ & $\dot{M}_{\rm{infall}}$ & $\rcirc$ & $T_{\rm{in}}$ & $p_{\text{in}}$   &    $\mbacc$   &   $\eta_{\rm vis}$  \\
        &   &   ($\mj$) &   ($\rj$) & &    & (\gcmc) &
        ($10^{-7}\frac{\mj}{\rm{yr}}$) &   ($\rpl$)   &   (K) &   (bar) &   ($10^{-7}\frac{\mj}{\rm{yr}}$)  &   \\
        \hline
        1 & super-Neptune    & 0.1 & 1 & $10^{-2}$ & $0.3\pi$ & $9.30\times10^{-14}$ & 0.5 & 4.94 & 1300     &   0.0233    &   0.246    &  0.841  \\
        2 & super-Neptune    & 0.1 & 1 & $10^{-2}$ & $0.25\pi$ & $3.86\times10^{-13}$ & 2.5 & 2.28 & 1300   &   0.0233    &   1.76  &   0.491     \\
        3 & super-Neptune    & 0.1 & 1 & $10^{-3}$ & $0.25\pi$ & $3.86\times10^{-13}$ & 2.5 & 2.28 & 1300   &   0.0233    &   1.91  &   0.487     \\
        4 & Saturn    & 0.3 & 1.5 & $10^{-2}$ & $0.3\pi$ & $9.24\times10^{-14}$ & 1.58 & 4.94 & 1400    & 0.0251  &   0.582 &   0.877      \\
        5 & Saturn    & 0.3 & 1.5 & $10^{-2}$ & $0.25\pi$ & $2.42\times10^{-13}$ & 5 & 2.28 & 1400    & 0.0251   &   3.38   &   0.578    \\
        6 & Saturn    & 0.3 & 1.5 & $10^{-3}$ & $0.25\pi$ & $2.42\times10^{-13}$ & 5 & 2.28 & 1400    & 0.0251   &   4.02   &   0.567    \\
        7 & Jupiter    & 1 & 1.8 & $10^{-2}$ & $0.3\pi$ & $3.80\times10^{-14}$ & 1.58 & 4.94 & 1400  & 0.0251     &   0.508 &   0.893    \\
        8 & Jupiter    & 1 & 1.8 & $10^{-2}$ & $0.25\pi$ & $1.01\times10^{-13}$ & 5 & 2.28 & 1400  & 0.0251  &   2.98   &   0.680    \\
        9 & Jupiter    & 1 & 1.8 & $10^{-3}$ & $0.25\pi$ & $1.01\times10^{-13}$ & 5 & 2.28 & 1400  & 0.0251  &   3.28   &   0.697 \\\hline\hline
    \end{tabular}
    \caption{Simulation parameters and bulk results on planetary accretion. From left to right: model ID, model description, planet mass, planet radius, Shakura-Sunyaev $\alpha$, $\theta_0$ (controlling the specific angular momentum of the infalling gas), $\rho_0$ (controlling the mass flux), calculated mass infall rate from one hemisphere, mean circularization radius of the infalling gas (Equation \ref{eqn:rcircmean}), inner-boundary temperature, inner-boundary pressure, measured accretion rate (one hemisphere), and viscous heating energy conversion fraction (defined in Section \ref{sec:conversion}).}
    \label{tab:bc}
\end{table*}

\begin{figure*}
    \centering
    \includegraphics[width=0.31\textwidth]{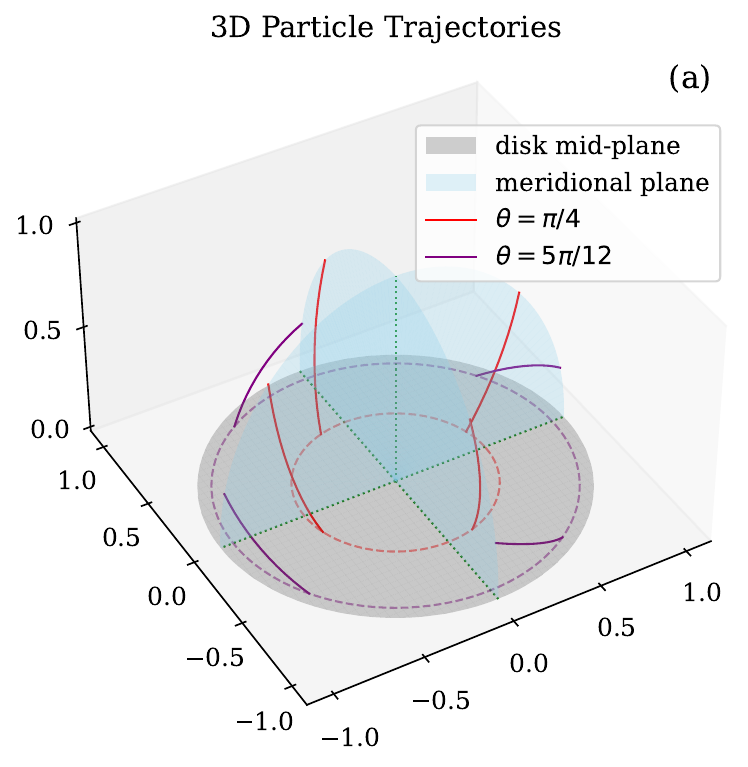}
    \includegraphics[width=0.335\textwidth]{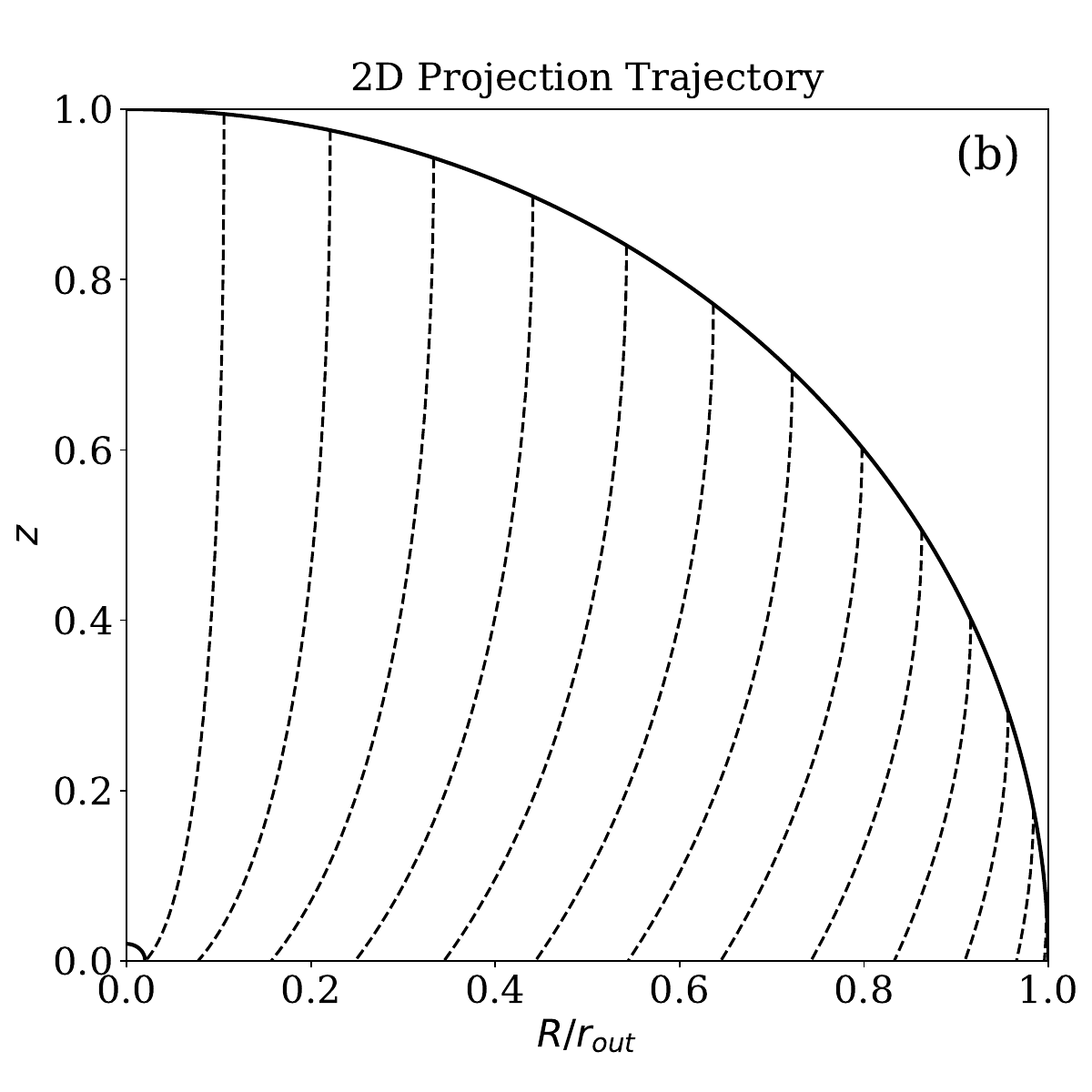}
    \includegraphics[width=0.335\textwidth]{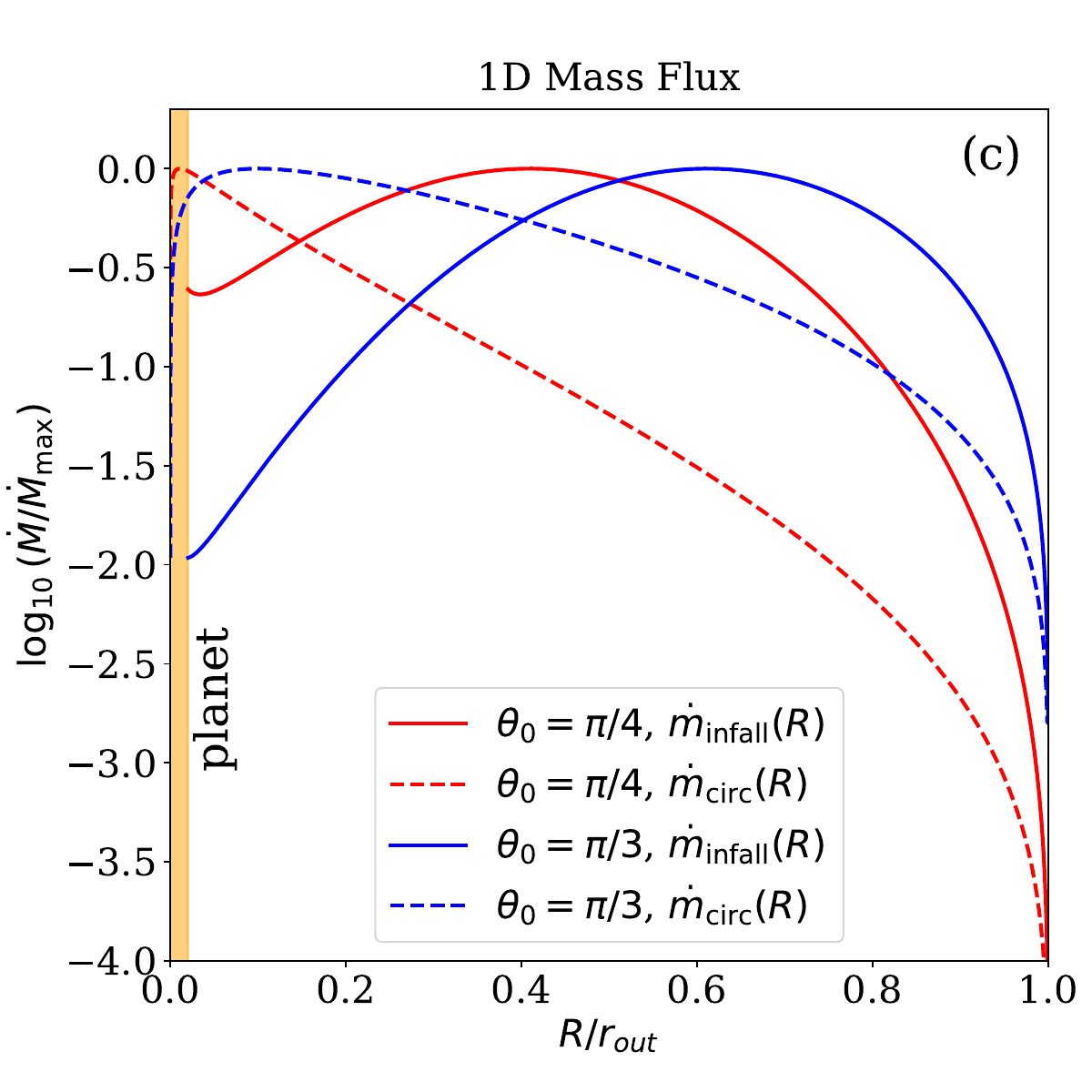}
    \caption{Panel (a): 3D ballistic trajectories corresponding to the boundary conditions described by Equations \ref{eqn:boundrho}--\ref{eqn:boundomega}. Test particles start from $\theta=\{\pi/4, 5\pi/12\}$ at azimuthal angles $\phi=\{0,\pi/2,\pi,3\pi/2,2\pi\}$, illustrated by the two meridional planes. The azimuthal symmetry indicates that particles starting from the same $\theta$ will hit the same radii on the midplane, shown by the dashed circles. Panel (b): ballistic trajectories from different $\theta$ projected onto the $r$-$\theta$ plane. Panel (c): midplane mass flux normalized to its maximum value. Solid lines indicate the mass flux when gas first hits the midplane; dashed lines indicate the mass flux if the gas immediately circularizes after it hits the midplane. Red and blue colors are associated with $\theta_{0}=\{0.25\pi, 0.3\pi\}$, respectively.}
    \label{fig:boundary}
\end{figure*}

We carry out a set of CPD simulations with different $\alpha$, planet masses $\mpl$, and radii $\rpl$ (see Table \ref{tab:bc}). The outer boundary mimics the infall of gas from the PPD by injecting cool gas with angular momentum along the $z$-axis. The gas subsequently falls towards the planet and forms a CPD. The inner boundary has a fixed temperature, density, rotation profile, and radius, designed to mimic the planetary atmosphere.

In this paper, we posit that the flow feeding the CPD arrives over a range of polar angles centered around some $\theta_0$ with non-zero angular momentum along the $z$-axis \citep{tanigawa2012,li2023,adams2025}, and has a vertically downward velocity profile in the meridional plane \citep{tanigawa2012}. The polar angle-dependent outer boundary conditions are
prescribed as:
\begin{eqnarray}
    \rho(\theta)&=&\rho_{0}\exp\bigg(-\frac{(\theta-\theta_{0})^{2}}{\Delta\theta^{2}}\bigg),     \label{eqn:boundrho}\\
    v_{r}(\theta)&=&-\sqrt{2G\mpl/\rout}\cos^{2}\theta,     \label{eqn:boundvr}\\
    v_{\theta}(\theta)&=&\sqrt{2G\mpl/\rout}\cos\theta\sin\theta,    \label{eqn:boundvtheta}\\
    \Omega(\theta)&=&\sqrt{G\mpl/\rout^{3}}\sin^{\xi}\theta,\label{eqn:boundomega}\\  
    \pdv{(r^{2}E_{r})}{r}&=&0  \label{eqn:bounder},
\end{eqnarray}
where we adopt $\xi=2.5$, $\Delta\theta=1/3$ rad, $\rout=50\,\rpl$ (much smaller than the typical Hill radius),
being fixed parameters in all our simulations, and the outer gas temperature is fixed to $T_{\text{out}}=20\,\mathrm{K}$. Equation \ref{eqn:bounder} allows the radiation to leave the computational domain in the free-streaming limit.

From Equations \ref{eqn:boundrho}, \ref{eqn:boundvr}, and \ref{eqn:boundomega}, we can define the mean circularization radius of the infalling gas by,
\begin{equation}\label{eqn:rcircmean}
    \rcirc=\frac{1}{G\mpl}\bigg(\frac{\int_{0}^{\pi/2}\rho v_{r}\Omega r^{2}\sin^{3}\theta d\theta}{\int_{0}^{\pi/2}\rho v_{r}\sin\theta d\theta}\bigg)^{2}\ ,
\end{equation}
which is mainly controlled by tuning the $\theta_0$ parameter. The mass inflow rate is mainly controlled by the $\rho_0$ parameter. These parameters are listed in Table \ref{tab:bc}. With the axisymmetric assumption, it is convenient to derive the ballistic trajectory of particles injected from any point on the outer boundary from Equations \ref{eqn:boundvr}, \ref{eqn:boundvtheta}, and \ref{eqn:boundomega} (see Appendix \ref{app:trajectory} for details). Figure \ref{fig:boundary}(a) and (b) show the ballistic trajectories in 3D starting from $\theta=\{\pi/4, 5\pi/12\}$ at different azimuthal angles, and trajectories starting from different $\theta$ projected onto the $r$-$\theta$ plane.
Once the particles reach the equator, we can numerically calculate the $R$-dependent mass flux in cylindrical coordinates $\dot{m}_{\text{infall}}(R)$ and the circularized mass flux $\dot{m}_{\text{circ}}(R)$ from Equation \ref{eqn:boundrho} (see Appendix \ref{app:trajectory}), shown in Figure \ref{fig:boundary}(c), where we define the circularized mass flux as the mass flux if the particle is immediately circularized.

For the inner boundary conditions, we adopt smaller planetary radii for small gas giant masses \citep{spiegel2012}. We use a fixed inner density $\rho_{\text{in}}=5\times10^{-7}$ \gcmc\ for all models and $T_{\text{in}}$ can be found in Table \ref{tab:bc}. The choice of $T_\text{in}$ is somewhat arbitrary but is based on our 1D accretion model calculation \citep{chen2022}.
The resulting pressure $p_\text{in}$ then
determines the ``depth" of the inner boundary in the atmosphere, although the relation between $p_\text{in}$, $\rpl$, and $\mpl$ is not necessarily self-consistent without a planetary interior and evolution model (but see \citet{zha2026}). This fixed-pressure inner boundary will allow gas to leave the computational domain and be accreted by the planet.
We further assume that the radiation energy flux at the inner boundary is zero and
the planet rotates differentially, following \citet{dong2021}:
\begin{equation}
    \pdv{E_{r}}{r}=0,\quad
    \Omega=\Omega_\text{in}\sin\theta,
\end{equation}
where $\Omega_\text{in}$ is set to 20\% of the planet's Keplerian frequency at $R=\rpl$. 

The computational domain is initially almost empty. To expedite the simulation, we artificially increase $\rho_0$ for a short period of time after the simulation begins. When the CPD has piled up sufficient material, we change $\rho_0$ back to its assumed value in Table \ref{tab:bc} and continue the simulation
until the accretion rate has approached a statistically steady state within $1.8\,\rpl$.
For reference, the effective viscous timescale can be estimated as
\begin{equation}
    \tvis\approx\frac{5.30\times10^{6}}{\alpha}\bigg(\frac{\mpl}{\mj}\frac{R}{10\,\rj}\bigg)^{0.5}\bigg(\frac{500\,\text{K}}{T}\bigg)\,\text{s}.
\end{equation}
Our simulations are run for a couple of $t_{\rm vis}$ when $\alpha=10^{-2}$ to reach steady state, but this is rarely achieved for $\alpha=10^{-3}$, or for very large and cool CPDs. However, our main purpose is to find quasi-steady states of CPDs that have reached their thermodynamic steady state (Section \ref{sec:conservation}). Because the viscous heating is dominated by the inner disk and the radiation transport is usually effective, the time it takes to reach the thermodynamic steady state is usually shorter than $t_{\rm vis}$.

\section{Results}\label{sec:results}

\subsection{Representative Quasi-steady State Solutions}\label{sec:quasisteadystate}

\begin{figure*}
    \centering
    \includegraphics[width=\textwidth]{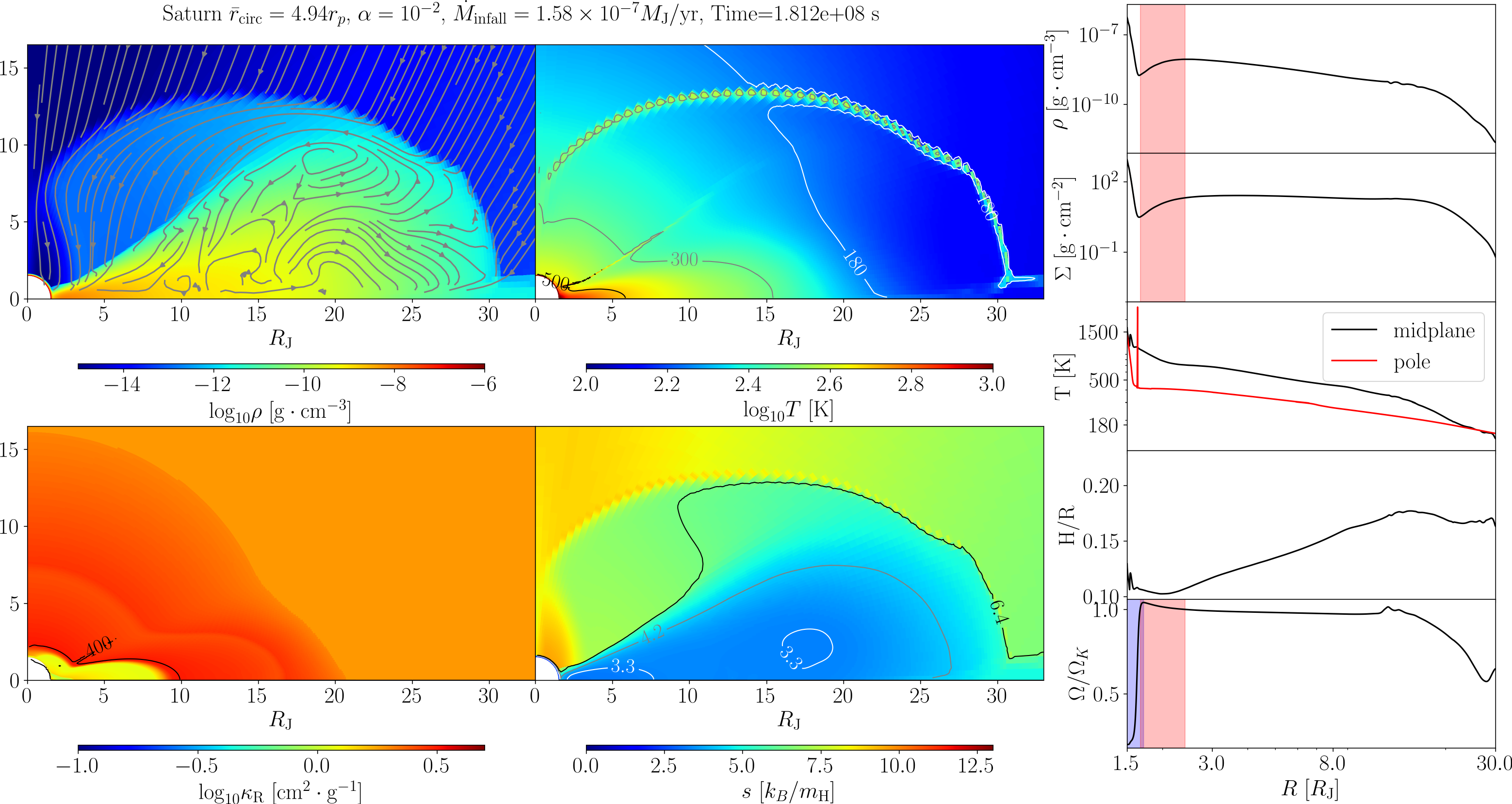}
    \caption{Quasi-steady-state solution of model 4. The left color panels show the profiles of density over-plotted with velocity streamlines, gas temperature with the $T=\{180,300,500\}\,\mathrm{K}$ contours, opacity with the $T=400\,\mathrm{K}$ contour, and entropy over-plotted with contours at $s=\{3.3,4.2,6.4\}k_{\rm B}\,m_{\rm H}^{-1}$. The right panels show the radial profiles of midplane density, surface density, temperature along the equator (black) and pole (red), midplane disk aspect ratio, and midplane rotational profile. The light blue and red shaded regions indicate the BL and super-Keplerian regions, respectively.
    }
    \label{fig:saturnbigcpd}
\end{figure*}

We choose two representative solutions (models 4 and 6) from the Saturn-mass planet models, shown in Figures \ref{fig:saturnbigcpd} and \ref{fig:saturnsmallcpd}. Results from the other simulations are qualitatively similar and are presented in Appendix \ref{app:qsssolutions}.

\begin{figure*}
    \centering
    \includegraphics[width=\textwidth]{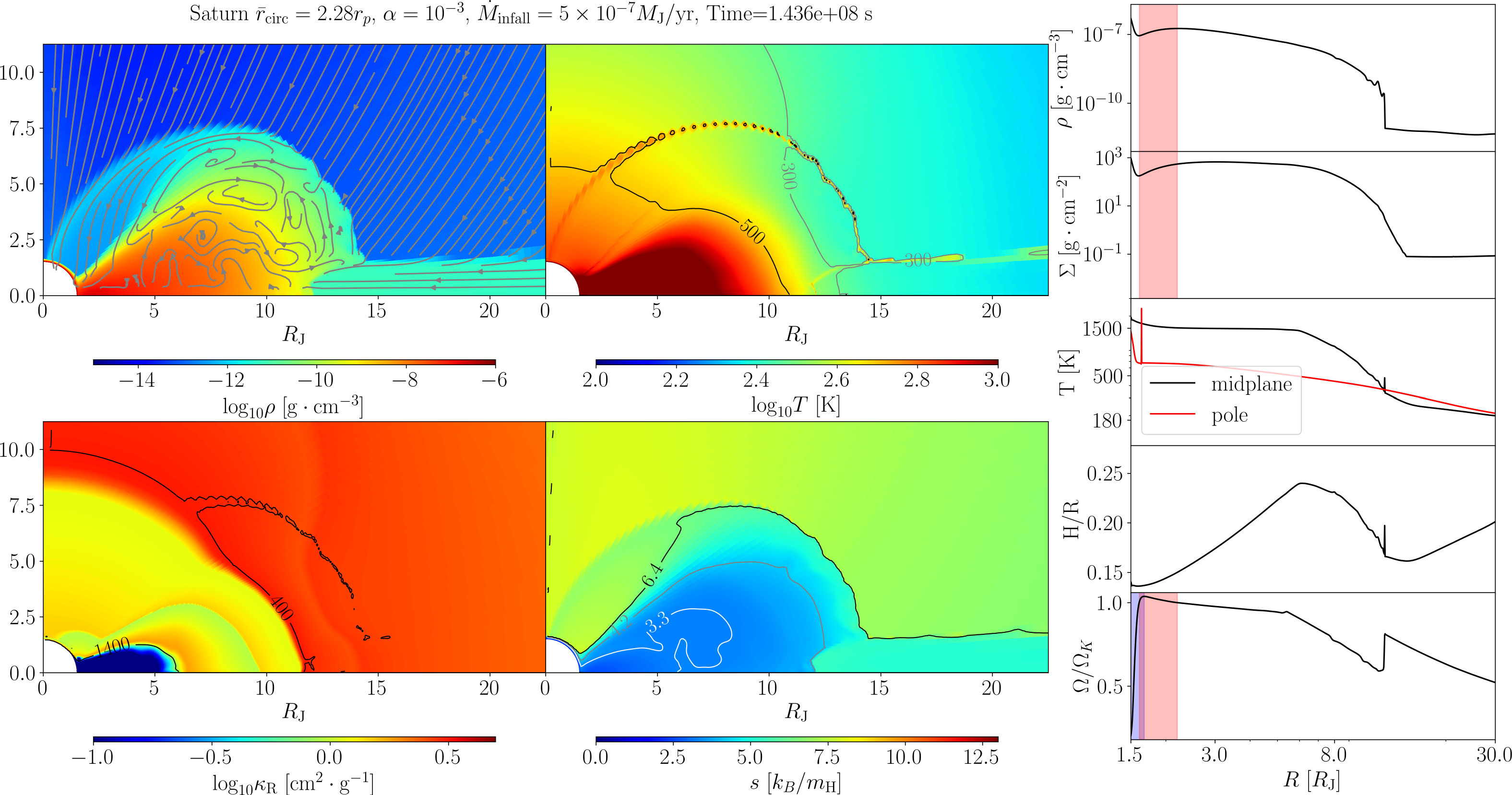}
    \caption{Quasi-steady-state solution of model 6 with the same layout as Figure \ref{fig:saturnbigcpd}. Because of the high temperature in the inner part of the CPD ($\lesssim6\,\rj$), dust sublimates and the opacity drops (blue color) by two orders of magnitude.}
    \label{fig:saturnsmallcpd}
\end{figure*}

\subsubsection{Disk overall structure}\label{sec:overallstruc}

In model 4, with a relatively large circularization radius $\rcirc=4.94\rpl$, the CPD extends to $20\,\rpl\approx30\,\rj$ at the end of the simulation---shown in the density plot---and would continue to extend if run longer. The streamlines show that some gas falls onto the outer CPD and is deflected to the inner CPD, which is discussed more in Section \ref{sec:accshock}. As has been widely identified in previous works \citep{fung2016,batygin2020}, the CPD exhibit a midplane decretion flow, where in our case spans from $15$ to $25\,\rj$. The overall disk temperature increases toward the planet from $\lesssim200\,\mathrm{K}$ to about $1000\,\mathrm{K}$. We overplot the $T=\{180,300,500\}\,\mathrm{K}$ contours on the temperature plot to delineate the low-temperature region; water ice may exist beyond the $T=180\,\mathrm{K}$ contour. The color change in the $\kr$ plot reflects the sublimation of some organic volatiles \citep{semenov2003}, highlighted by the $T=400\,\mathrm{K}$ contour. Silicate dust remains intact in Figure \ref{fig:saturnbigcpd} because $T$ has not reached the silicate sublimation temperature ($\sim1500\,\mathrm{K}$); therefore, $\kr\in(1,4)$ \cmsg\ throughout the computational domain. The entropy $s$ is discussed in Section \ref{sec:entropy} in detail.

The right panels of Figure \ref{fig:saturnbigcpd} show the midplane radial profiles of $\rho$, surface density $\Sigma$, $T$, the disk aspect ratio $H/R$, and the Keplerian ratio $\Omega/\Omega_{\rm K}$.
The aspect ratio, here between $0.1$ and $0.2$, is generally an increasing function of radius before the disk is truncated, and is discussed further in Section \ref{sec:cpecpd}.
The red line in the temperature panel shows $T$ along the pole. A Zel'dovich spike emerges at the planetary surface, produced by an accretion shock that directly heats the surface as infalling gas impacts it, consistent with 1D models \citep{marleau2019,chen2022}. Given our prescribed gas infall profile, this accretion shock carries only a tiny fraction ($<1\%$) of the inflowing gas. By contrast, there is no accretion shock in the BL; rather, the CPD temperature is enhanced by viscous dissipation there.

In the Keplerian ratio plot, the BL is clearly visible. Within the BL (shaded light blue), $\Omega$ decreases from super-Keplerian to the planetary spin rate $\Omega_{\text{in}}$. A dip in $\rho$ can be understood from the simplified momentum equation
\begin{equation}\label{eqn:cylindricaldisk}
    v_{r}\pdv{v_{r}}{R}+\frac{1}{\rho}\pdv{p}{R}+(\Omega^{2}_{K}-\Omega^{2})R=0.
\end{equation}
In this region, since $|v_{r}|<0.1c_{T}$ and $v_r$ is smooth, the first term is negligible compared with the others. In the red region, $\Omega^{2}_{K}-\Omega^{2}<0$, indicating $\partial p/\partial R>0$ in the BL; combined with $\partial T/\partial R<0$, this implies $\partial\rho/\partial R>0$, consistent with the dip in the density profile. Because the accretion rate is almost constant in the inner region of the CPDs (Section \ref{sec:conservation}), the density dip also means an inward acceleration of $|v_{r}|$ by several times near the BL, driven by the viscous torque.

Figure \ref{fig:saturnsmallcpd} shows the solution of model 6. Its CPD is much smaller than that in model 4, owing to its smaller $\rcirc$, and has a much higher density because of its higher $\dot{M}_{\text{infall}}$ and smaller $\alpha$. Its decreation flow is around $6$-$10\,\rj$ near the midplane. The column density $\Sigma=\int_{z}\rho dz$ is correspondingly higher, since $\dot{M}_{\text{infall}}\propto\alpha\Sigma$. The increased density produces a larger optical depth and, ultimately, weaker cooling, so $T$ inside the CPD rises. Once $T\gtrsim1500\,\mathrm{K}$, silicate dust sublimates and the very inner region of the CPD becomes transparent, as seen in the $\kr$ color plot. This transparent region is nearly isothermal owing to efficient radial radiation transport, evident in the midplane $T$ profile.

\subsubsection{Entropy}\label{sec:entropy}

We use entropy $s$ to highlight the heating and cooling processes at the quasi-steady state. Because we use $\gamma$-law gas, we calculate the entropy per hydrogen mass ($k_{\rm B}\,m_{\rm H}^{-1}$) by,
\begin{equation}
    s=\frac{1}{\mu}\left(\frac{\gamma}{\gamma-1}\ln{\frac{T}{T_0}}-\ln{\frac{p}{\rm{p_0}}}\right),
\end{equation}
where $T_0=1000$\,K and $p_0=1$\,bar define our zero-entropy reference level. At this $(T_{0},p_{0})$ state, $\rm{H}_{2}$ is purely molecular and remains so for $T<1600$\,K (i.e., throughout the CPDs). A realistic equation of state (EoS) predicts $s_{\rm real}=8.8\,k_{\rm B}\,m_{\rm H}^{-1}$ at this reference state \citep[see Equation 29a of][]{marleau2017}. Because $s_{\rm real}>11\,k_{\rm B}\,m_{\rm H}^{-1}$ is typically considered ``hot'' in the context of planet formation \citep{spiegel2012}, and assuming no hydrogen dissociation \citep{chen2022}, any value of $s>2.2\,k_{\rm B}\,m_{\rm H}^{-1}$ relative to our zero point is considered ``hot'' in this work.

When the infalling gas approaches the CPD, a shock forms at the disk atmosphere and outer edge, which heats the gas to roughly $500\,\mathrm{K}$ ($s>7\,k_{\rm B}\,m_{\rm H}^{-1}$); it then cools to around $200\,\mathrm{K}$ ($s\sim6.4\,k_{\rm B}\,m_{\rm H}^{-1}$) in the post-shock region due to optically thin radiation transport.
The entropy plot shows that strong cooling at the shock front causes an abrupt entropy drop in the post-shock region, which continues as the infalling gas settles toward the disk. As accretion proceeds, heating and cooling are approximately in balance; as the gas is further compressed and cooled, entropy continues to decrease toward the inner CPD. However, entropy slightly increases near the BL, as shown by the $s=3.3\,k_{\rm B}\,m_{\rm H}^{-1}$ contour detaching from the planet in Figure \ref{fig:saturnbigcpd}---a result of the strong viscous heating occurring there. Across all nine simulations, the gas accreted through the BL is strongly heated, reaching $s>2.2\,k_{\rm B}\,m_{\rm H}^{-1}$ (equivalent to $s_{\rm real}>11\,k_{\rm B}\,m_{\rm H}^{-1}$ for a realistic gas).

\subsection{Energy Conservation and Conversion}\label{sec:conservation}

During accretion, gravitational potential energy is converted into internal energy via viscous heating, compressional work, or shocks, and is ultimately dissipated as radiation. Understanding the complex thermodynamics of CPDs therefore requires a careful accounting of their energetics. We address this by analyzing the detailed pathways of energy conservation and conversion.

\subsubsection{Energy conservation}
In a quasi-steady state, energy conservation can be characterized by the balance between the energy inflows and outflows. Adding Equations \ref{eqn:totalE} and \ref{eqn:rader} yields the total energy conservation equation,
\begin{equation}\label{eqn:energycon}
    \pdv{(E+E_{r})}{t}+\nabla\cdot{[(E+p)\vel+\F-\Omega\boldsymbol{\tau}]}=\rho\vel\cdot\mathbf{g}.
\end{equation}

In the quasi-steady state, the first term on the LHS of Equation \ref{eqn:energycon} may be dropped, yielding
\begin{equation}
    \nabla\cdot{[(E+p)\vel+\F-\Omega\boldsymbol{\tau}]}\approx-\rho \vel\cdot\nabla\Phi.\label{eqn:energyconserv}
\end{equation}
where $\Phi=-G\mpl/r$ is the gravitational potential. In the quasi-steady state limit,
\begin{equation}
    \rho\vel\cdot\nabla\Phi\approx\nabla\cdot(\rho\vel\Phi).\label{eqn:gravcon}
\end{equation}

Substituting Equation \ref{eqn:gravcon} into Equation \ref{eqn:energyconserv}, integrating over the hemisphere, and applying Gauss's law transforms the volumetric conservation law into an energy flux balance on the spherical shell of radius $r$,
\begin{equation}
    \int_{S}[(E+p)v_{r}+F_{r}-\Omega\tau_{r\phi}]dS\approx\int_{S}\frac{-\rho v_{r}G\mpl}{r}dS+C,\label{eqn:conserve}
\end{equation}
where $F_{r}=\F\cdot\hat{r}$ is the radial radiation flux and $dS=2\pi r^{2}\sin\theta d\theta$. We gauge energy conservation by checking whether Equation \ref{eqn:conserve} holds at different radii.
We define the total and gravitational energy fluxes as
\begin{align}
    f_{\text{total}}(r)&=\int_{S}[(E+p)v_{r}+F_{r}-\Omega\tau_{r\phi}]dS,  \label{eqn:sumflux}\\
    f_{\Phi}(r)&=-\frac{G\mflux(r)\mpl}{r}+C, \label{eqn:phi}
\end{align}
where $\mflux(r)=\int_{S}\rho v_{r}dS$ is the radius-dependent mass flux, influenced by mass accumulation at different radii and viscous disk expansion. Once the disk reaches a quasi-steady state, we approximate $\mflux(r)$ by a constant accretion rate $\dot{\bar{m}}_{\rm acc}$.

\begin{figure}
    \centering
    \includegraphics[width=\columnwidth]{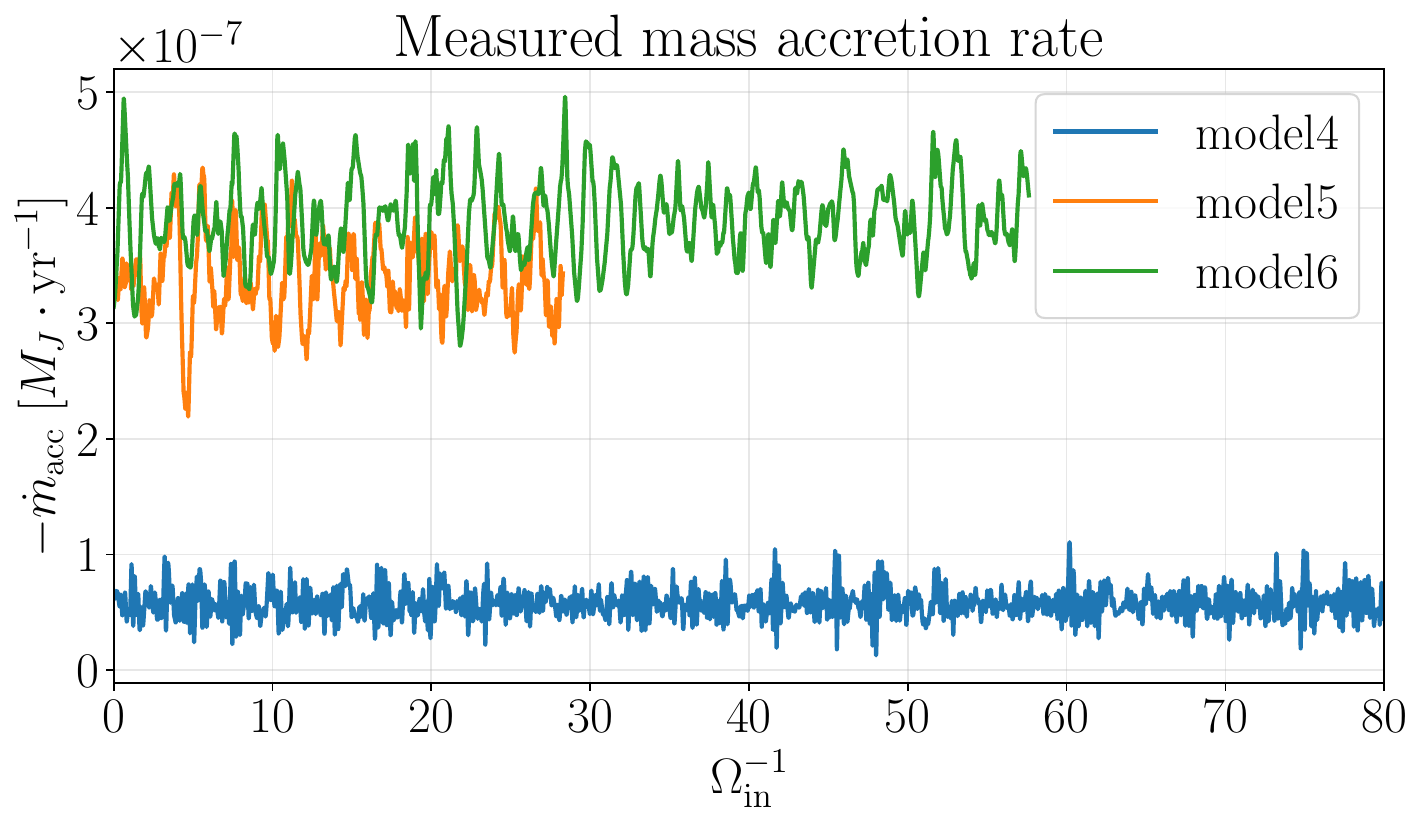}
    \includegraphics[width=\columnwidth]{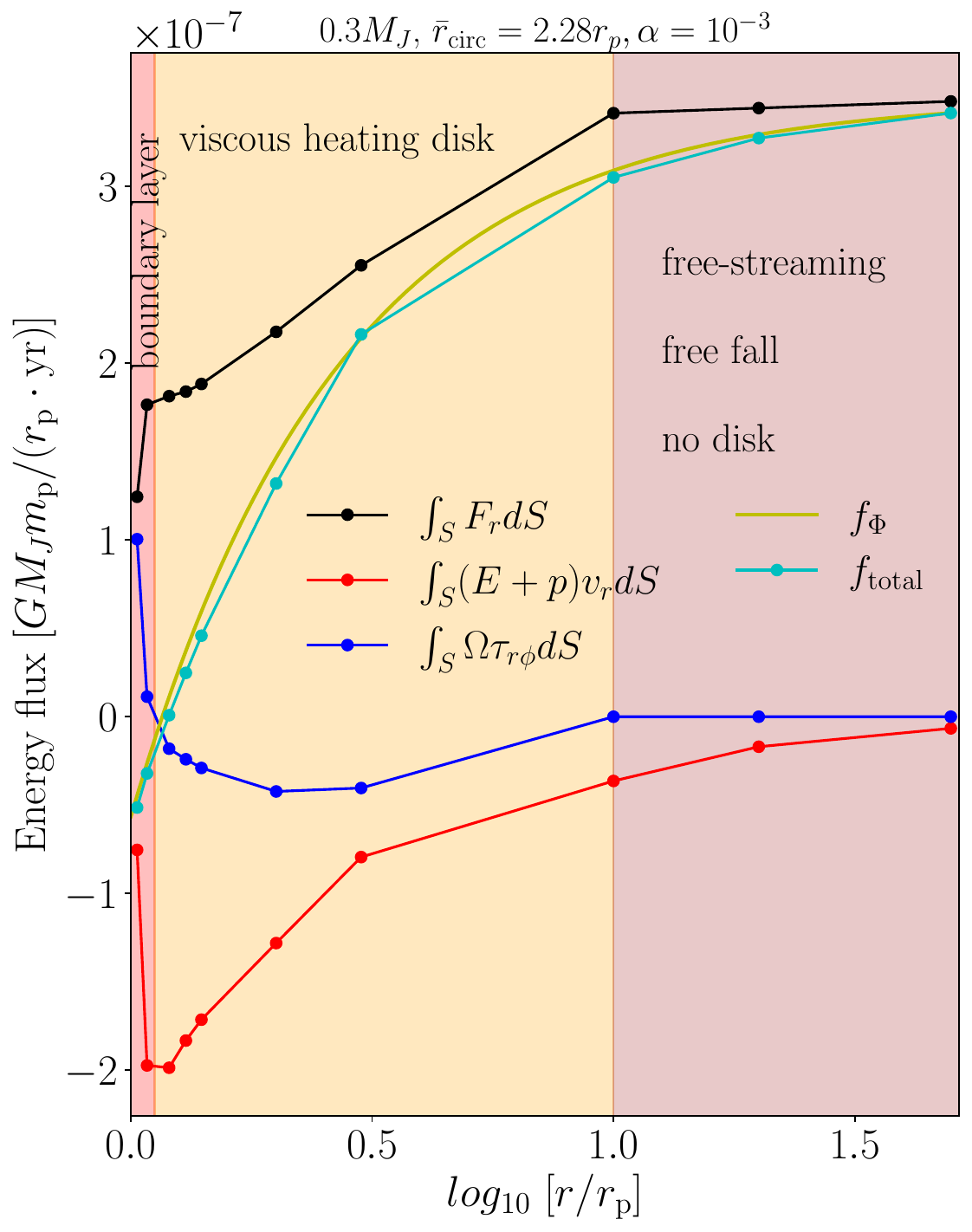}
    \caption{Upper panel: the integrated mass flux history at $1.2\,\rpl$ for models 4, 5, and 6. Time is measured in units of $\Omega_{\rm{in}}^{-1}$, where $\Omega_{\rm{in}}$ is 20\% of the planet's Keplerian frequency. Bottom panel: the integrated energy fluxes at sampled radii of model 6. $f_{\rm{total}}$ and $f_{\Phi}$ are defined in Equations \ref{eqn:sumflux} and \ref{eqn:phi}, respectively.}
    \label{fig:energyflux}
\end{figure}

We estimate $\macc$ statistically in three steps. First, we sample $\macc$ at $r=\{1.2, 1.5, 1.8\}\,\rpl$ at high frequency and time-average. Second, if the averaged values at all three radii agree to within $5\%$ (except for model 9, which is computationally expensive), we regard the inner disk as having reached a quasi-steady state and $\macc$ as well-defined. Third, we average $\macc$ over the three radii to obtain $\dot{\bar{m}}_{\rm acc}$. The three representative accretion histories shown in the upper panel of Figure \ref{fig:energyflux} all display statistical convergence. The constant $C$ is fixed once $f_{\text{total}}(r)$ is known at certain radii.

To compute $f_{\text{total}}$, we integrate $(E+p)v_{r}$, $F_{r}$, and $\Omega\tau_{r\phi}$ over 10 shells in the range $r\in[1.08,50]\,\rpl$ and time-average. All fluxes are normalized by $G\mj\mpl/(\rpl\cdot\rm{yr})$, so that their numerical values correspond to an ``accretion luminosity in $\mj/\rm{yr}$". The sum of the three fluxes gives $f_{\text{total}}$, and we fix $C$ by equating $f_{\text{total}}(\rout)$ and $f_{\Phi}(\rout)$. The fluxes of model 6 are shown in the bottom panel of Figure \ref{fig:energyflux}; results for the other runs appear in Appendix \ref{app:fluxes}. The closeness $f_{\text{total}}\approx f_{\Phi}$ indicates good energy conservation within the computational domain. The slightly non-zero $f_{\Phi}(\rpl)<0$ implies that the inner boundary acts as a mild energy sink, possibly owing to the subsonic inner boundary conditions.

\subsubsection{Physical processes behind energy conversion}\label{sec:conversion}
We highlight three distinct regions of energy conversion, visible in the bottom panel of Figure \ref{fig:energyflux}.
\begin{enumerate}[leftmargin=15pt]
    \item Freefall advection: The light brown region is the disk-free zone ($r>10\,\rpl$), characterized by $\Omega\tau_{r\phi}\approx0$. Here, gas free-falls toward the CPD and planet while the radiation is free-streaming, as shown by the nearly constant $\int_{S}F_{r}dS$.
    \item Viscous disk heating: Once the gas hits the CPD, the free fall ends. The light orange region marks the bulk disk, characterized by $\Omega\tau_{r\phi}<0$.
    The negative $\tau$ means the outer disk spins down the inner disk. The radiative luminosity (black line) rises gradually with $r$ here, as viscous heat and gas internal energy are transported outward by radiation.
    \item BL heating: The light red region is the BL, characterized by $\Omega\tau_{r\phi}>0$. The positive $\tau$ means the outer disk spins up the inner disk. The highly localized viscous heating in the BL produces the jump in the radiative luminosity (black line) across this region, accounting for nearly half of the total luminosity (the inner boundary has zero radiative luminosity).
\end{enumerate}

To further analyze the energetics, we define the fraction of viscous heating relative to the total accretion luminosity as
\begin{equation}
    \eta_{\rm vis}=\frac{\rpl\int_{V}\bar{\dot{\Lambda}} dV}{G\mpl\mbacc},
\end{equation}
where $\bar{\dot{\Lambda}}$ is the averaged viscous heating rate. The results for each model are listed in Table \ref{tab:bc}. Viscous heating clearly dominates in CPDs with large $\rcirc$. When $\rcirc$ is small, CPDs morphologically resemble compact envelopes, and compressional work converts a large fraction of gravitational energy into internal energy.

In summary, energy conversion shifts from compressional work in compact envelopes to viscous dissipation in extended circumplanetary disks. Within these extended disks, this viscous heating is highly localized in the boundary layer and generates nearly half of the total accretion luminosity.

\subsection{The Thermal Structure of CPDs}\label{sec:cpecpd}

\begin{figure}
    \centering
    \includegraphics[width=\columnwidth]{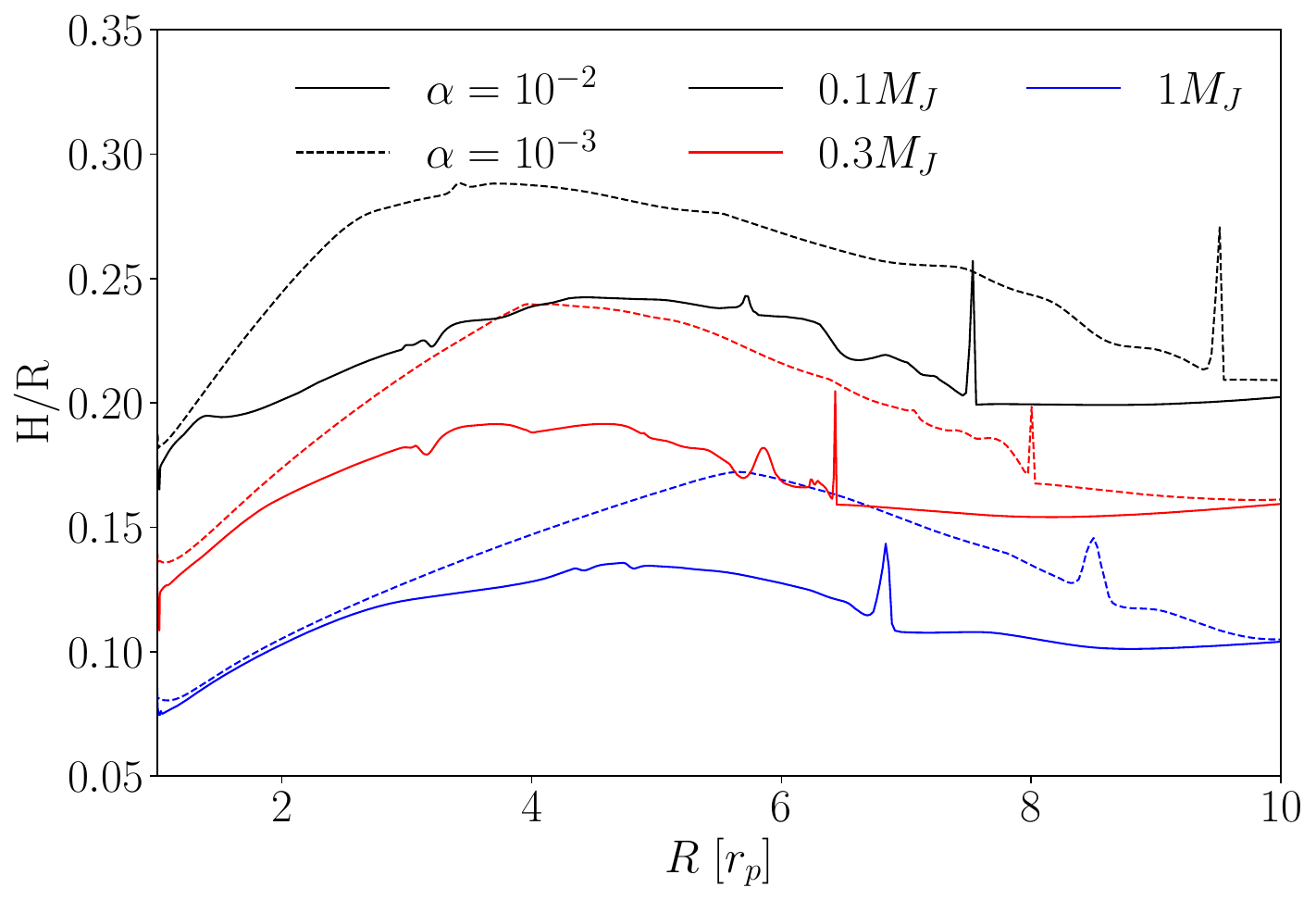}
    \includegraphics[width=\columnwidth]{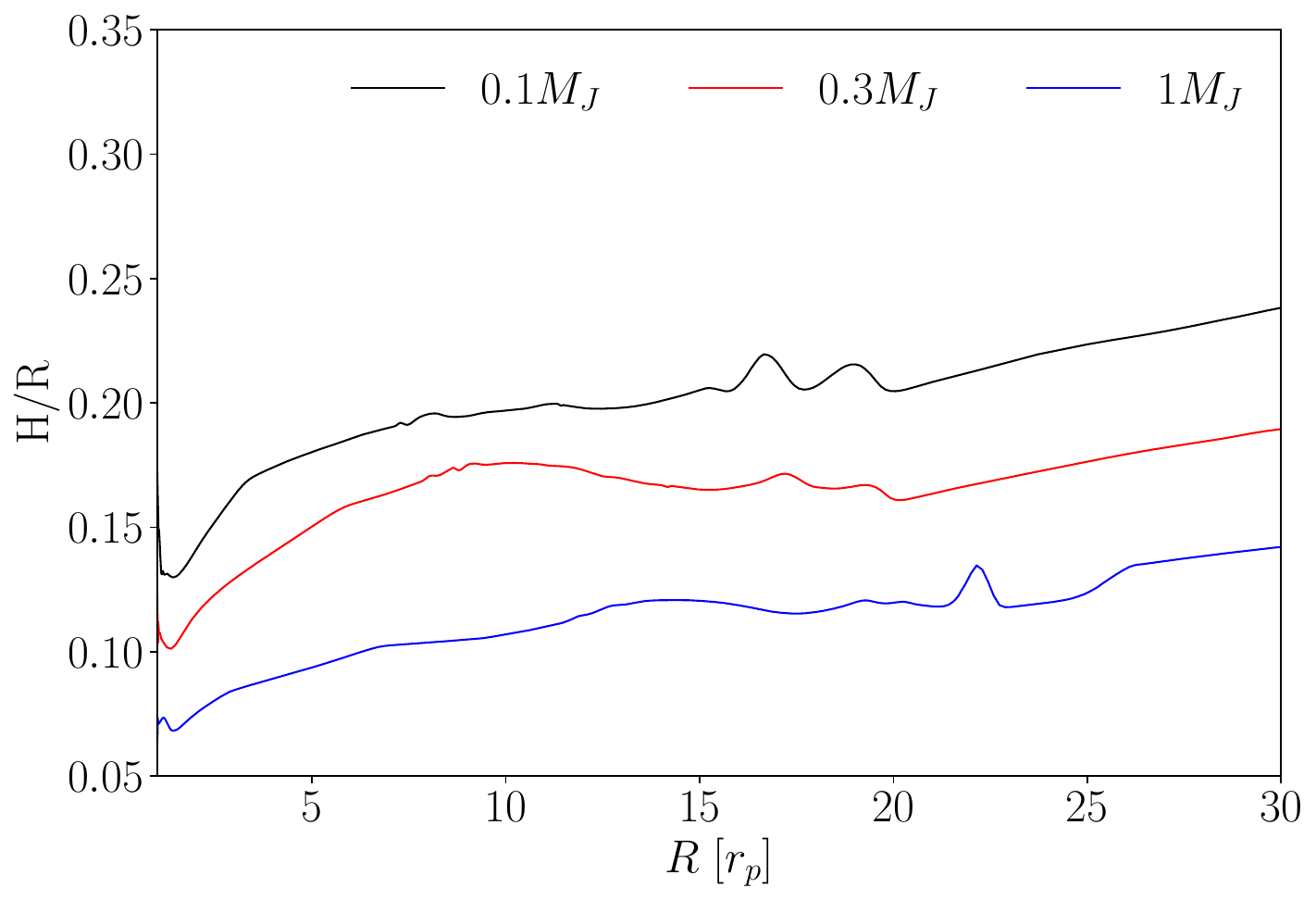}
    \caption{Upper panel: $H/R$ of models with $\rcirc=2.28\,\rpl$. Solid lines have $\alpha=10^{-2}$ and dashed lines have $\alpha=10^{-3}$. The spikes/bumps in the outer region are shocks induced by the infalling gas and indicate the disk truncation radii. Bottom panel: $H/R$ of models with $\rcirc=4.94\,\rpl$ and $\alpha=10^{-2}$.}
    \label{fig:hr}
\end{figure}

\subsubsection{Disk aspect ratio}

A disk's thermal structure is usually characterized by its aspect ratio $H/R$. In our model, the CPD thermal structure is strongly influenced by the BL and viscous heating (Section \ref{sec:conversion}); radiation transport is another important process shaping the midplane $T$. In a classic viscous Keplerian thin disk with torque-free inner boundary condition, constant mass accretion rate $\dot{M}$, $\alpha$-viscosity, and opacity $\kr$, the midplane temperature $T_{c}$ follows from balancing viscous heating against vertical radiation transport, while the surface density $\Sigma$ follows from balancing the viscous torque against angular-momentum transport:
\begin{eqnarray}\label{eqn:tc}
    T_{c}&\approx&\left( \frac{27 k_{\rm B} \kr \alpha \Sigma^2}{64 \mu m_H \sigma_{\rm{SB}}} \sqrt{\frac{G\mpl}{R^3}} \right)^{1/3},  \\
    \Sigma&=&\frac{\mu m_{\rm H}\dot{M}}{3\pi k_{\rm B}\alpha T_{c}}\sqrt{\frac{G\mpl}{R^3}}\bigg(1-\sqrt{\frac{\rin}{R}}\bigg). \label{eqn:sigma}
\end{eqnarray}
Equation \ref{eqn:tc} holds when the disk is optically thick \citep[$\tau=\kr\Sigma\gg 1$,][]{shakura1973,frank1992}. Physically, $\kr$ can change drastically as the temperature crosses critical values, fundamentally altering the energy transport and temperature profile.

Substituting $\Sigma$ into Equation \ref{eqn:tc} and solving for $T_c$ yields the analytic formula
\begin{align}
    &\frac{T_c}{\rm{K}} \approx 1.11 \times 10^4 \left( \frac{\alpha}{0.01} \right)^{-0.2} \left( \frac{\kr}{1\,\rm{cm}^{2}\,\rm{g}^{-1}} \right)^{0.2} \left( \frac{\mpl}{\mj} \right)^{0.3}  \nonumber\\
    &\left( \frac{\dot{M}}{10^{-6}\,\mj\,\rm{yr}^{-1}} \right)^{0.4}  \left( \frac{R}{\rj} \right)^{-0.9}\bigg(1-\sqrt{\frac{\rin}{R}}\bigg)^{0.4}, \label{eqn:tc2}
\end{align}
and substituting $T_c$ into $H/R$ gives
\begin{align}
    &\frac{H}{R}=0.151\times\left(\frac{\alpha}{0.01}\right)^{-0.1}\left(\frac{\kr}{1\,\rm{cm}^2\,\rm{g}^{-1}}\right)^{0.1} \left(\frac{\mpl}{\mj}\right)^{-0.35}  \nonumber\\
    &\left(\frac{\dot{M}}{10^{-6}\,\mj\,\rm{yr}^{-1}}\right)^{0.2} \left(\frac{R}{\rj}\right)^{0.05}\bigg(1-\sqrt{\frac{\rin}{R}}\bigg)^{0.2}.  \label{eqn:correlation}
\end{align}

Figure \ref{fig:hr} shows the aspect ratio across all models. Substituting the simulation parameters into this equation (recall that $\mbacc$ in Table \ref{tab:bc} is measured for one hemisphere) allows a direct comparison between the 2D simulations and the analytic model. In general, the analytic trends carry over to the 2D simulations: $H/R$ is negatively correlated with $\alpha$ and $\mpl$. The analytic formula nonetheless predicts a higher $H/R$ by a factor of $\approx1.2$ at large radii, even accounting for BL heating in the 2D simulations; this implies a lower $T_c$ in the 2D runs, which may reflect more efficient energy dissipation.

\subsubsection{Boundary layer heating}

\begin{figure}
    \centering
    \includegraphics[width=\columnwidth]{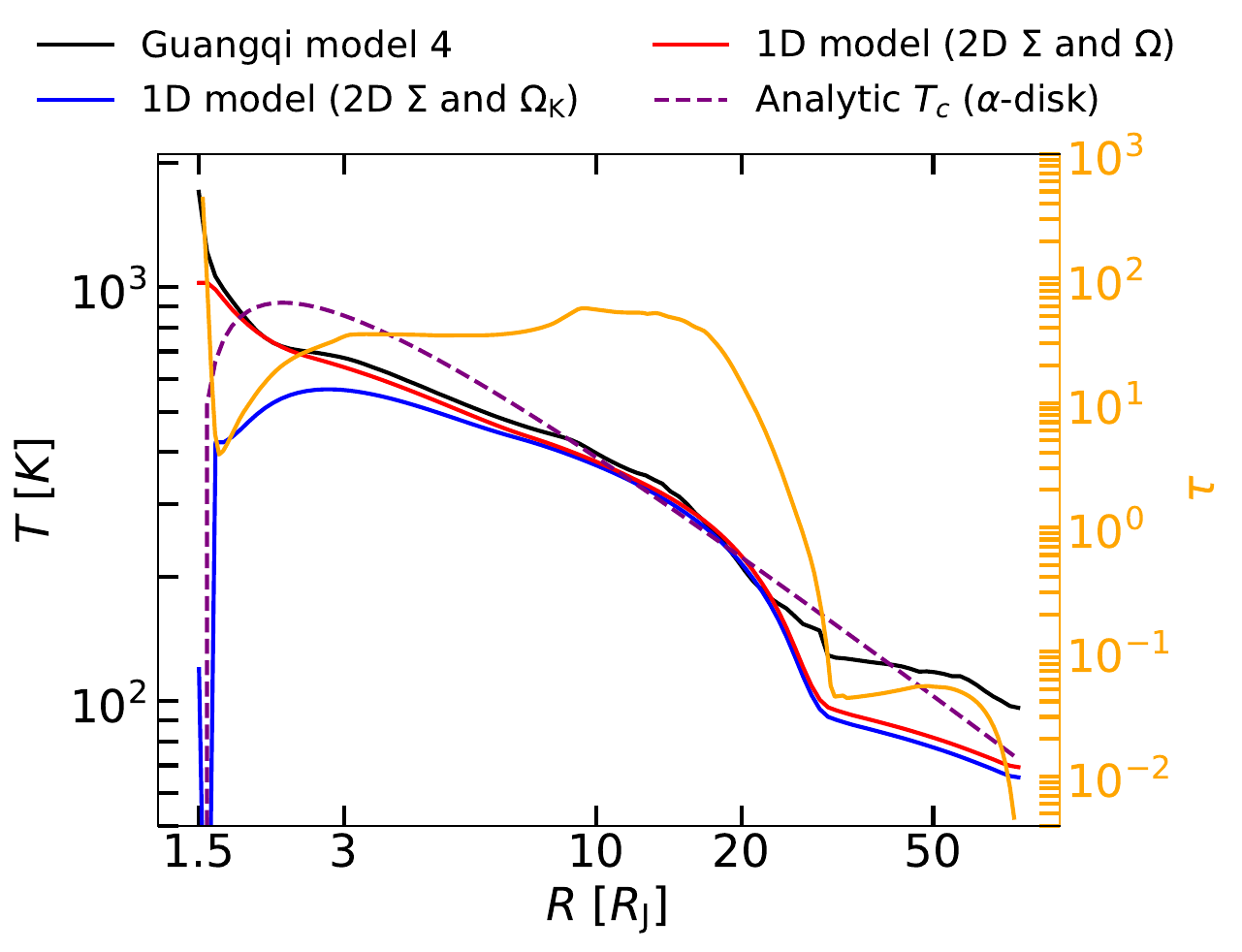}
    \caption{Left $y$-axis: midplane temperature profiles of model 4 (typical), comparing the analytic, 1D, and 2D results. The black, blue, red, and purple-dashed lines are the results from {\tt Guangqi} in this work, the thin-disk model using only $\Sigma$ from the 2D result, the thin-disk model using both $\Sigma$ and $\Omega$ from the 2D result, and the analytic formula with $\dot{M}=2\mbacc$ and $\kr=2\,\rm{cm}^{2}\,\rm{g}^{-1}$, respectively. Right $y$-axis: the optical depth $\tau$ of model 4.}
    \label{fig:cpd1d2d}
\end{figure}

To understand the physical origin of this discrepancy and enable a clear comparison, we employ a modified 1D viscous thin-disk model \citep{ni2026}. This model computes local viscous heating from $\Omega$ and $\Sigma$, coupling it to radiation transport to obtain the equilibrium temperature.

A fundamental difference between the 2D model and the analytic model is the BL. The BL is sub-Keplerian and can contribute significantly to disk heating \citep{kley1989}, potentially accounting for up to half of the total luminosity, so the Keplerian assumption underlying the classic models becomes inappropriate for CPD BLs. We examine two scenarios with the 1D model: (1) $\Sigma$ from the 2D simulation with Keplerian $\Omega_{\rm K}$, which has no BL; (2) both $\Sigma$ and $\Omega$ from the 2D simulation, which has a self-consistent BL structure and heating. The results, together with the 2D simulation and the analytic formula from Equation \ref{eqn:tc2}, are shown in Figure \ref{fig:cpd1d2d}. The close agreement between the red line and the 2D results within the disk region ($1.6<R<25\,\rj$) indicates that a 1D model captures the essential thermodynamics---including viscous heating and radiation transport---provided it incorporates the correct hydrodynamic structure. The innermost part of the 2D simulation reflects the temperature profile of the planetary atmosphere, not the CPD. The analytic result underestimates $T_c$ near the BL, owing to the torque-free inner boundary. Meanwhile, it overestimates $T_c$ around $3\rj$ when $\tau\lesssim40$, and is more accurate around $15\rj$ when $\tau\gtrsim60$. The over-estimation of $T_c$ by the analytic formula may stem from the Eddington approximation and optically thick assumption, which becomes less accurate at intermediate to low optical depth.

These thermodynamic limitations are especially critical when the CPD density is high, and when BL heating, viscous dissipation, compressional work, and radiative cooling collectively dictate the CPD temperature profile. Compared to the classic thin-disk, our 2D model more self-consistently resolves these thermodynamic processes, providing a new framework for studying non-magnetized CPDs.

\subsection{Accretion shocks}\label{sec:accshock}

\begin{figure}
    \centering
    \includegraphics[width=\columnwidth]{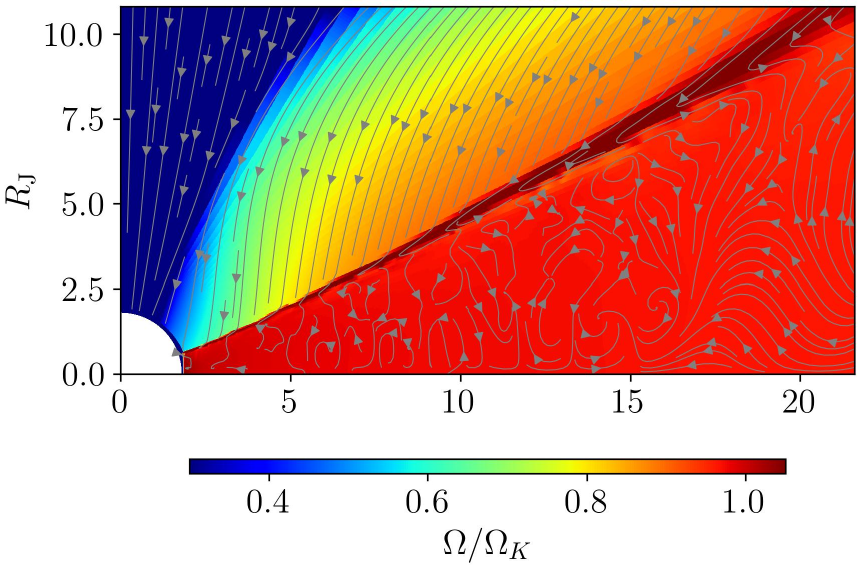}
    \caption{Ratio $\Omega/\Omega_{\rm K}$ overplotted with streamlines for model 9. The low-angular-momentum gas slides directly onto the planetary surface, whereas gas with higher specific angular momentum hits the CPD first, forming a super-Keplerian layer.}
    \label{fig:omegaratio}
\end{figure}

\begin{figure*}
    \centering
    \includegraphics[width=\textwidth]{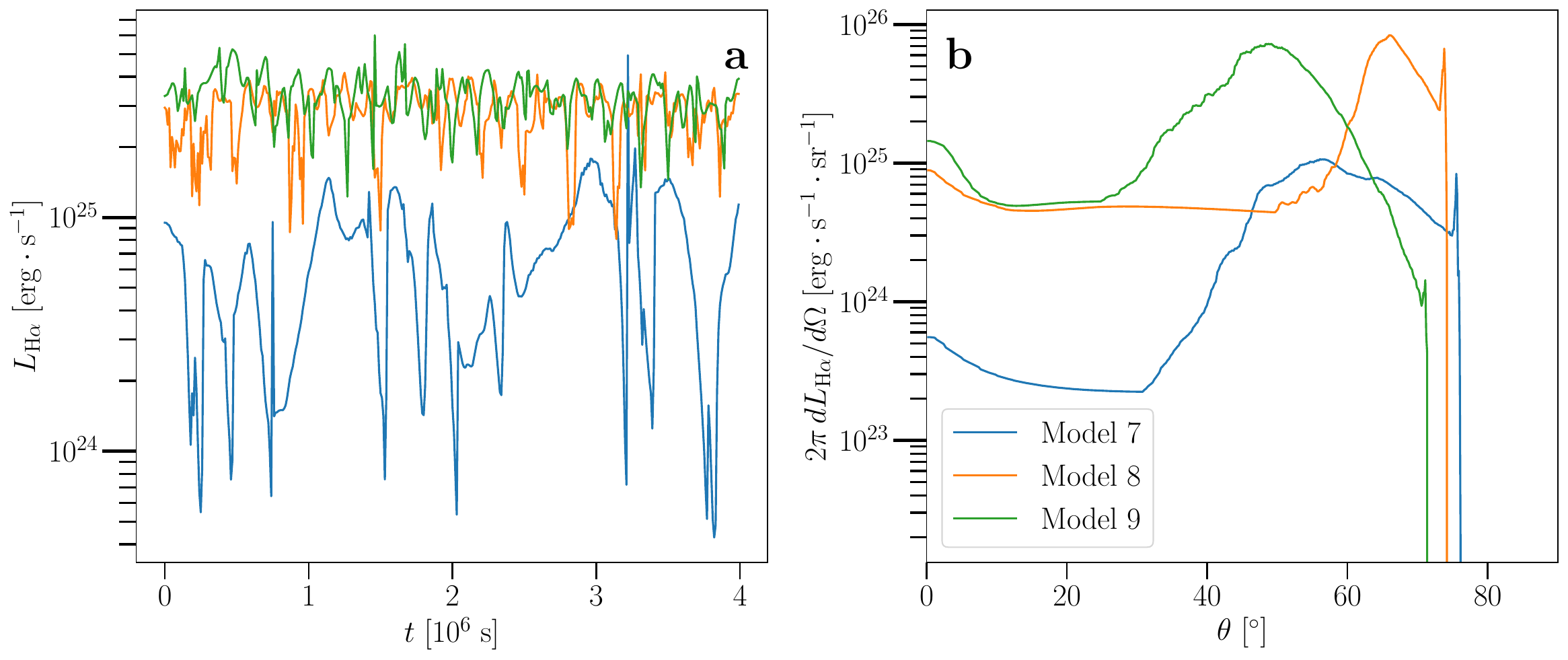}
    \caption{Left: the estimated $L_{\rm{H}{\alpha}}$ of models 7, 8, and 9 over the whole sphere. Right: the angular dependence of $L_{\rm{H}{\alpha}}$.}
    \label{fig:lhalpha}
\end{figure*}

Consistent with \citet{takasao2021}, a supersonic accretion flow exists at high elevation angles above the quasi-steady CPD. Streamline plots (Figures \ref{fig:omegaratio} and \ref{fig:jupitersmallcpd}) indicate that this flow originates from the high-altitude outer regions of the slightly flared CPD, intercepting infalling gas from the outer boundary. Gas with low specific angular momentum slides along the CPD surface and falls directly onto the planet, forming a shock, whereas gas with higher angular momentum along the $z$-axis falls onto the CPD.

We calculate the mass accretion rates through the shocks for the Jupiter-mass models by integrating the supersonic flux over the hemisphere and averaging over $\gtrsim0.12\,{\rm yr}$. Doubling this value to account for the full sphere yields $\dot{M}_{\rm shock}=\{4.45\times10^{-9},\ 1.38\times10^{-8},\ 1.66\times10^{-8}\}\,\mj\,{\rm yr}^{-1}$ for model $\{7, 8, 9\}$. This constitutes only $2\%$--$5\%$ of the total accretion rate in the simulations (Table \ref{tab:bc}). We estimate the total H$\alpha$ luminosity ($L_{\rm{H}{\alpha}}$) by converting $1\%$ of the kinetic energy of the supersonic flow where $|v_{r}|>v_{\mathrm{crit}}=30\,\mathrm{km}\,\mathrm{s}^{-1}$ \citep{aoyama2018}, i.e.,
\begin{equation}\label{eqn:lalpha}
    \frac{L_{\rm{H}\alpha}}{\rm{erg}\,\rm{s}^{-1}} \approx 10^{27} \cdot \frac{\mpl}{3\mj} \cdot \frac{\dot{M}_{\rm shock}(v_{r}<v_{\rm crit})}{5.5\times10^{-8}\,\mj\,\text{yr}^{-1}} \cdot \frac{1.8\,\rj}{\rpl},
\end{equation}
and show the result in Figure \ref{fig:lhalpha}a. We also calculate the time-averaged angular dependent $L_{\rm{H}\alpha}$ and show in Figure \ref{fig:lhalpha}b. Within the supersonic flow, due to the low free fall speed in all these three simulations, only $40$-$44\%$ of the supersonic flow satisfies $|v_{r}|>v_{\rm crit}$, giving rise to $L_{\rm{H}{\alpha}}\lesssim4\times10^{25}\,\rm{erg}\,\rm{s}^{-1}$. In contrast, PDS 70c \citep{zhou2025} has $L_{\rm{H}{\alpha}}\gtrsim10^{27}\,\rm{erg}\,\rm{s}^{-1}$. Due to the very high computational demand for the high mass planet simulations, the most massive planet in our simulation is 1$\mj$. If we run simulations for massive planets ($\mpl\gtrsim3\mj$), the supersonic flow that hits the planetary surface could be energetic enough to produce observable H$\alpha$ emission when $\dot{M}_{\rm shock}\gtrsim5.5\times10^{-8}\,\mj\,\text{yr}^{-1}$ \citep[Equation \ref{eqn:lalpha};][]{aoyama2019}, while even more massive ones ($\mpl\gtrsim5\mj$) may produce H$\alpha$ emission from shocks at the CPD surface as infalling gas hits the disk \citep{marleau2023}.


\section{Discussion}\label{sec:discussion}

\subsection{Implications for satellite formation}

The disk temperature naturally decreases outward. In systems with $\dot{M}_{\mathrm{infall}} < 10^{-6} \mj/\mathrm{yr}$ (integrated over the full sphere), the $T=180\,\mathrm{K}$ isotherm---beyond which water ice can survive---varies from $10$ to $35\,\rj$ across super-Neptune, Saturn, and Jupiter-mass protoplanets (excluding model 9, which has not reached a quasi-steady state; see Figures \ref{fig:saturnbigcpd}, \ref{fig:saturnsmallcpd}, and \ref{fig:neptunebigcpd}--\ref{fig:jupitersmallcpd}). The water-ice line therefore lies somewhat beyond this isotherm. The large radial distance of this isotherm for Saturn and Jupiter may imply that the Galilean satellites accreted ice pebbles at large radii before migrating inward during the late-stage CPD evolution \citep{fujii2026}, or that they formed in the late stage.

In this work, we considered BL accretion from the CPD, relevant when the surface density of the CPD is high ($\Sigma>1\,$\gcmc) \citep{fujii2014}, and the planetesimal formation may be in progress. In contrast, when most of the CPD mass is depleted, magnetospheric accretion \citep{takasao2022,zhu2024} may operate in magnetized protoplanets harboring low-density CPDs, where magnetic torque truncates the inner disk. Such cavities are likely crucial for moon migration \citep{shibaike2019,fujii2020,batygin2025,fujii2026}. Nevertheless, the timing and radius of disk truncation remain highly uncertain, primarily because the protoplanets' magnetic field strengths \citep{elias2025,zha2026} and CPD ionization fractions \citep{fujii2014} are poorly constrained.

\subsection{Caveats and future work}

We modeled CPD accretion through a BL with a constant effective $\alpha$-viscosity \citep{popham1993,dong2021} throughout the computational domain. The physical origin of viscosity within the BL, however, likely differs from that in the outer CPD: angular momentum may be transported non-locally by acoustic waves \citep{belyaev2013} or Rossby waves \citep{fu2023}, phenomena whose accurate modeling demands high azimuthal resolution---beyond the scope of this work.

Finally, dust opacity can fundamentally reshape CPD structure (Section \ref{sec:cpecpd}) and thermodynamic state, not only through sublimation (Section \ref{sec:quasisteadystate}) but also through coagulation and fragmentation \citep{yang2026}. Because dust evolution is highly sensitive to the thermal environment, a coupled treatment of thermodynamics and dust physics is needed for a more realistic picture of CPDs.

\section{Conclusions}\label{sec:conclusion}

We have presented a viscous, axisymmetric RHD model of CPDs developed using the {\tt Guangqi} code. The model employs the Shakura-Sunyaev $\alpha$-viscosity prescription \citep{shakura1973}, the flux-limited diffusion (FLD) approximation for radiation transport \citep{levermore1981}, and fixed gas and dust opacity tables \citep{semenov2003,malygin2014}. We evolve angular momentum using the PSAMA method \citep{chen2026} on a geometrically stretched $(r,\theta)$ grid with static mesh refinement (Section \ref{sec:computationaldomain}).

To validate the model, we performed nine simulations (Table \ref{tab:bc}), eight of which are evolved to a quasi-steady state while ensuring energy conservation (Section \ref{sec:conservation}). Infalling gas from the PPD first impacts the CPD's extended envelope before settling onto the disk. Driven by viscous torques, a fraction of this gas is accreted through the CPD and BL, while the rest remains in the CPD and decreates (Section \ref{sec:quasisteadystate}). Our main findings are summarized as follows:
\begin{enumerate}[leftmargin=15pt]
    \item CPD temperature increases inward, and for actively accreting CPDs, the water iceline can be pushed beyond tens of $\rj$ (Figure \ref{fig:saturnbigcpd}, e.g., $\mpl>0.3\,\mj$, $\mbacc>1.16\times10^{-7}\,\mj\,\rm{yr}^{-1}$). The dust can sublimate in the innermost CPD, which becomes optically thin and isothermal. (Figure \ref{fig:saturnsmallcpd}).
    \item The entropy of the CPD and the gas accreted via the BL falls within the ``hot'' regime ($
    s_{\rm real}>11\,k_{\rm{B}}\,m_{\rm{H}}^{-1}$) across the nine simulations. This suggests the ``hot-start'' phase extends through the BL accretion stage (Sections \ref{sec:entropy} and \ref{sec:conversion}).
    \item The classic thin-disk model with a torque-free inner boundary cannot reproduce the BL structure seen in our 2D RHD simulations, and therefore underestimate the temperature near the BL; while overestimates temperatures at intermediate optical depths (Section \ref{sec:cpecpd}).
    \item We identify a supersonic flow with low angular momentum, carrying a few percent of the total accretion rate, that plunges directly onto the planet from high above the quasi-steady CPD. For massive planets ($\mpl\gtrsim3\,\mj$), the resulting shock could generate observable H$\alpha$ emission (Section \ref{sec:accshock}).
\end{enumerate}

In summary, this RHD model provides a solid foundation for studying critical aspects of planet formation, including chemical evolution \citep{oberg2023,cridland2025}, spin and angular momentum regulation \citep{batygin2018,dong2021,batygin2025}, the reproduction of CPD SEDs \citep{zhu2015a,choksi2025,shibaike2026,sun2026,taylor2026}, and H$\alpha$ emission from accretion shocks \citep{takasao2021,zhou2025}. It also supplies accurate initial conditions for modeling dust evolution and satellite formation \citep{fujii2020,huang2022,shulik2025,fujii2026}. Because the timescales of these processes typically exceed current simulation durations, future work will pursue long-term CPD simulations spanning $10^{3}$--$10^{5}$ years \citep{yang2027}.

\software{{\tt Guangqi} \citep{chen2026}, {\tt Matplotlib} \citep{hunter2007}, {\tt Petsc} \citep{petsc-efficient,petsc-user-ref}}

\begin{acknowledgments}
The authors thank Philip Armitage, Douglas Lin, Ruobing Dong, Yuhiko Aoyama, Kengo Tomida, Yuri Fujii, Nienke van der Marel, and Hidekazu Tanaka for helpful discussion. 
This work is supported by the National Science Foundation of China under grants No. 12233004, 12103028, 12325304, 12342501, Tsinghua University Dushi Program and Tsinghua University Initiative Scientific Research Program No. 20233080026. The Center for High Performance Computing at Tsinghua University and the National Supercomputer Center in Tianjin provided the computational resources.
\end{acknowledgments}

\bibliography{cpd}
\bibliographystyle{aasjournalv7}

\appendix

\section{Ballistic trajectory}\label{app:trajectory}

In the outer region of our computational domain, where the density is relatively small and the thermal effect has little impact, the collapse flow, which tends to travel under the principle of least action,
can be well described by a ballistic model. Here, we give the solution under the outer boundary conditions described in Section \ref{sec:ibc}.

The classical central-force problem in the inverse square force field gives three constants of motion: total energy $E_{\rm orb}$, angular momentum vector $\mathbf{L}$, and Laplace-Runge-Lenz vector $\mathbf{A}$. The elliptical orbital parameters are completely determined by the following,
\begin{equation}
    e = \sqrt{1+2E_{\rm orb}L^2/(\mpl^2G^2)}
\end{equation}
\begin{equation}
    a = -\mpl G/(2E_{\rm orb})
\end{equation}
\begin{equation}
    b = \sqrt{-L^{2}/(2E_{\rm orb})}
\end{equation}
where $e$ is the eccentricity; $a$ and $b$ are the lengths of the semi-major and semi-minor axes. Moreover, $\mathbf{A}$ gives the vector from the center of force to the orbit's perihelion; its expression is,
\begin{equation}
\mathbf{A}=\mathbf{v}\times\mathbf{L} -\mpl G \cdot\mathbf{\hat{r}}
\end{equation}

Moreover, $-\mathbf{A}$ gives the direction of the long axis of the orbit, denoted by the unit vector $\mathbf{\hat{a}}$, and $\mathbf{L}\times \mathbf{A}$ gives the direction of the short axis, denoted by the unit vector $\mathbf{\hat{b}}$. Then, the elliptical orbit can be represented by the parametric equation,
\begin{equation}
    \mathbf{r} = (ae+a\cos\psi)\mathbf{\hat{a}} - (b\sin\psi)\mathbf{\hat{b}}
\end{equation}
where $\psi$ is the eccentric anomaly, ranging from $0$ to $2\pi$ for a closed orbit. The first intersection of $\mathbf{r}$ with the mid-plane is the infall radius, denoted by $R_{\text{infall}}$. It has a complex functional relation with the outer boundary condition and injection angle $\theta$, 
which we denote as $f_{\theta}(\theta)=R_{\text{infall}}$. 
After intersecting the disk mid-plane, the vertical velocity of the particles
is expected to damp out, and they find a new elliptic orbit in the mid-plane, as shown in Figure \ref{fig:boundary}.

In the right panel of Figure \ref{fig:boundary}, we further numerically calculate the cylindrical-radius-$R$-dependent mass flux $\dot{m}_{\text{infall}}(R)$ from the outer boundary conditions by
\begin{equation}
    \int_{R_1}^{R_2}R\dot{m}_{\text{infall}}(R)dR=-\rout^{2}\int_{\theta_{1}}^{\theta_{2}}\rho v_{r}\sin\theta d\theta,
\end{equation}
where $\rho, v_{r}$ are given by Equation \ref{eqn:boundrho} and \ref{eqn:boundvr}, $f_{\theta}(\theta_{1})=R_1$, and $f_{\theta}(\theta_{2})=R_2$.

The circularization radius at a specific angle has a simpler relation with the outer boundary condition,
\begin{equation}\label{eqn:rcirc}
    r_{\text{circ}}(\theta) = \frac{\Omega^{2}(\rout\sin\theta)^{4}}{G\mpl}=g_{\theta}(\theta)
\end{equation}
where $\Omega$ is given by Equation \ref{eqn:boundomega}. The circularized mass flux can be calculated in a similar way as $\dot{m}_{\text{infall}}(R)$,
\begin{equation}
    \int_{R_1}^{R_2}R\dot{m}_{\text{circ}}(R)dR=-\rout^{2}\int_{\theta_{1}}^{\theta_{2}}\rho v_{r}\sin\theta d\theta,
\end{equation}
where $g_{\theta}(\theta_{1})=R_1$, and $g_{\theta}(\theta_{2})=R_2$.

\section{Simulation results of super-Neptune and Jupiter}\label{app:qsssolutions}

\begin{figure*}
    \centering
    \includegraphics[width=\textwidth]{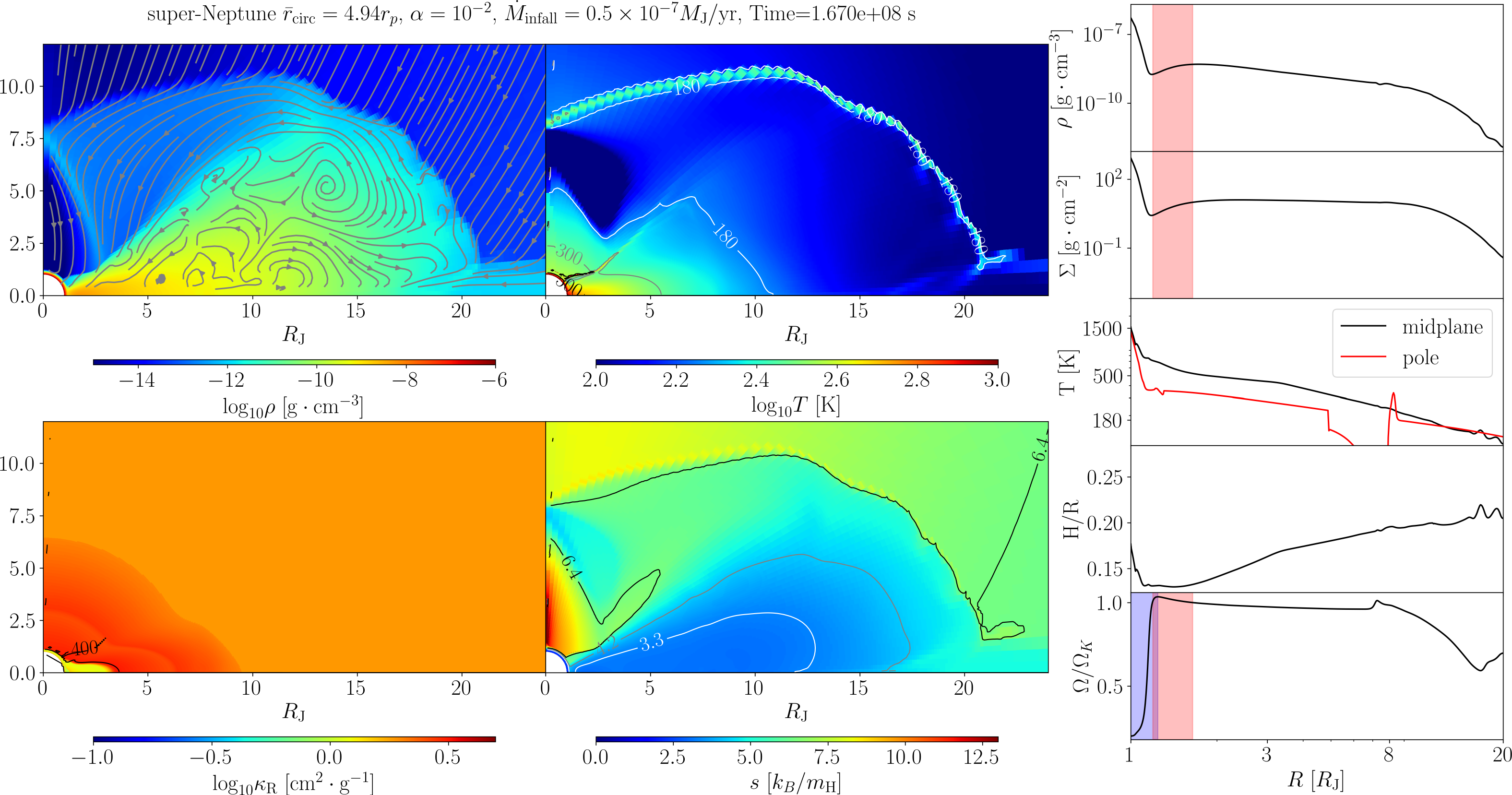}
    \caption{Quasi-steady-state solution of model 1, with the same layout as Figure \ref{fig:saturnbigcpd}. The decreation flow spans from $11$ to $16\,\rj$ near the midplane.}
    \label{fig:neptunebigcpd}
\end{figure*}

\begin{figure*}
    \centering
    \includegraphics[width=\textwidth]{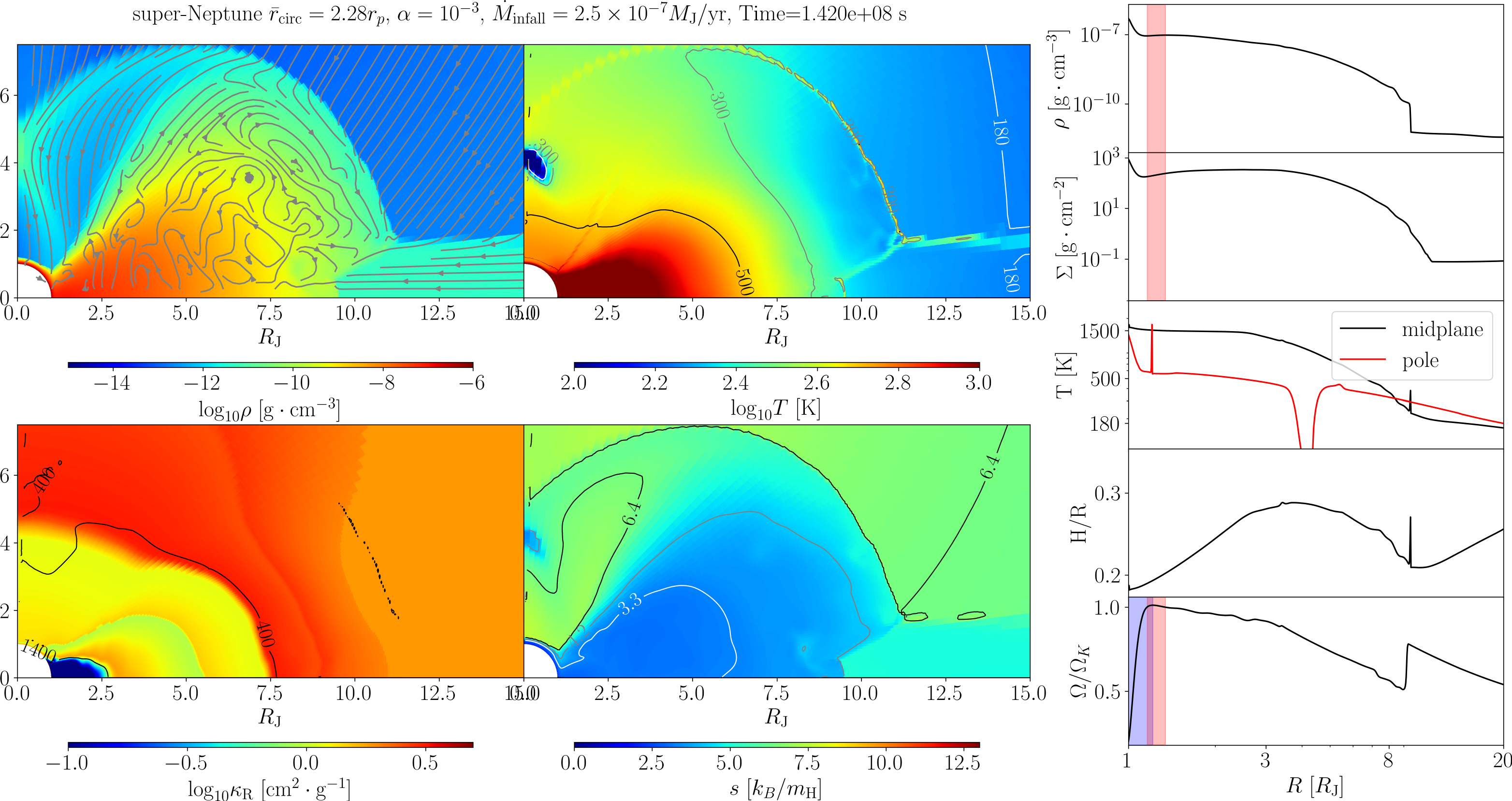}
    \caption{Quasi-steady-state solution of model 3, with the same layout as Figure \ref{fig:saturnbigcpd}. The streamlines of the decreation flow are not significant in this simulation.}
    \label{fig:neptunesmallcpd}
\end{figure*}

\begin{figure*}
    \centering
    \includegraphics[width=\textwidth]{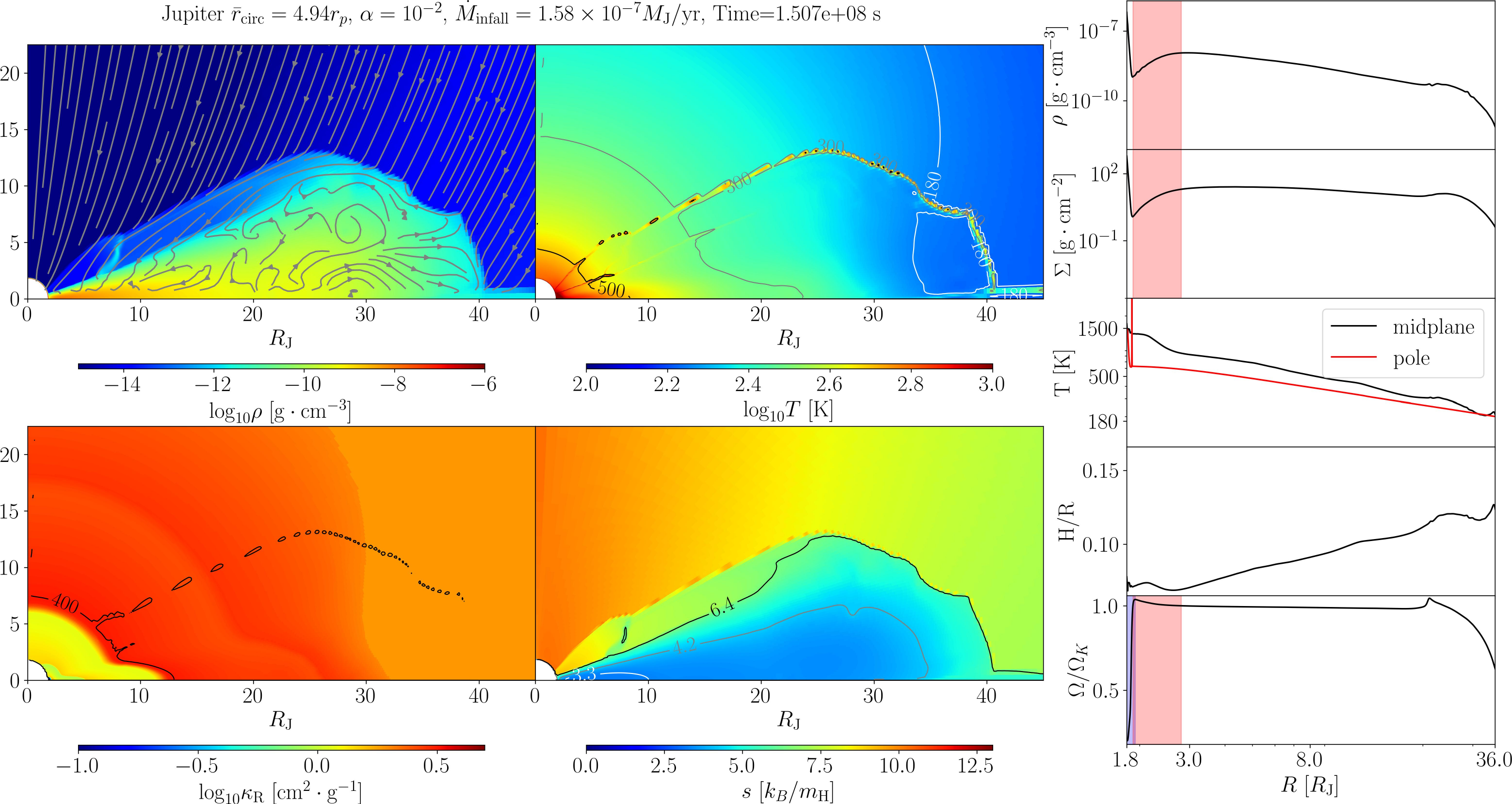}
    \caption{Quasi-steady-state solution of model 7, with the same layout as Figure \ref{fig:saturnbigcpd}. The decreation flow spans from $25$ to $40\,\rj$.}
    \label{fig:jupiterbigcpd}
\end{figure*}

\begin{figure*}
    \centering
    \includegraphics[width=\textwidth]{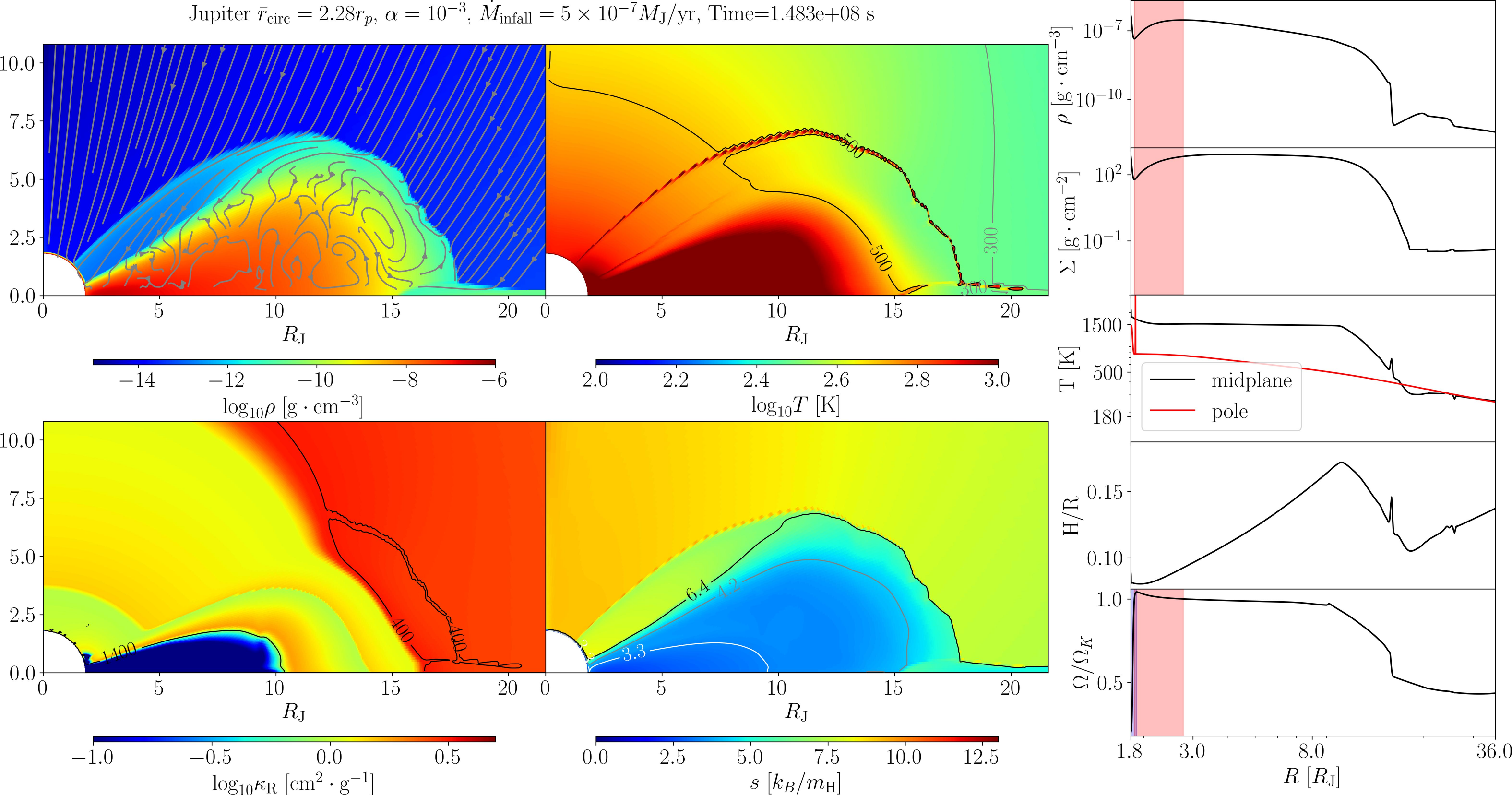}
    \caption{Final simulation state of model 9, with the same layout as Figure \ref{fig:saturnbigcpd}. The decreation flow spans from $10$ to $15\,\rj$. The $180\,\mathrm{K}$ temperature contour extends out to $60\,\rj$; we note that this simulation has not yet reached its quasi-steady state, owing to the very large $t_{\rm{vis}}$.}
    \label{fig:jupitersmallcpd}
\end{figure*}

Figures \ref{fig:neptunebigcpd}--\ref{fig:jupitersmallcpd} show the quasi-steady-state solutions for models 1, 3, and 7, as well as the final simulation state for model 9, following a similar pattern to Figure \ref{fig:saturnbigcpd}. Models 2, 5, and 8 yield results similar to models 3, 6, and 9, but with lower temperatures and densities.

The temperature anomaly near the poles in some simulations likely arises from the hydro solver struggling with converging flows in that region. However, because the local density and mass flux are negligible compared to the CPD, this numerical artifact does not affect our primary conclusions. We anticipate resolving this issue in future models.


\section{Energy fluxes and conservation}\label{app:fluxes}

\begin{figure*}
    \centering
    \includegraphics[width=\textwidth]{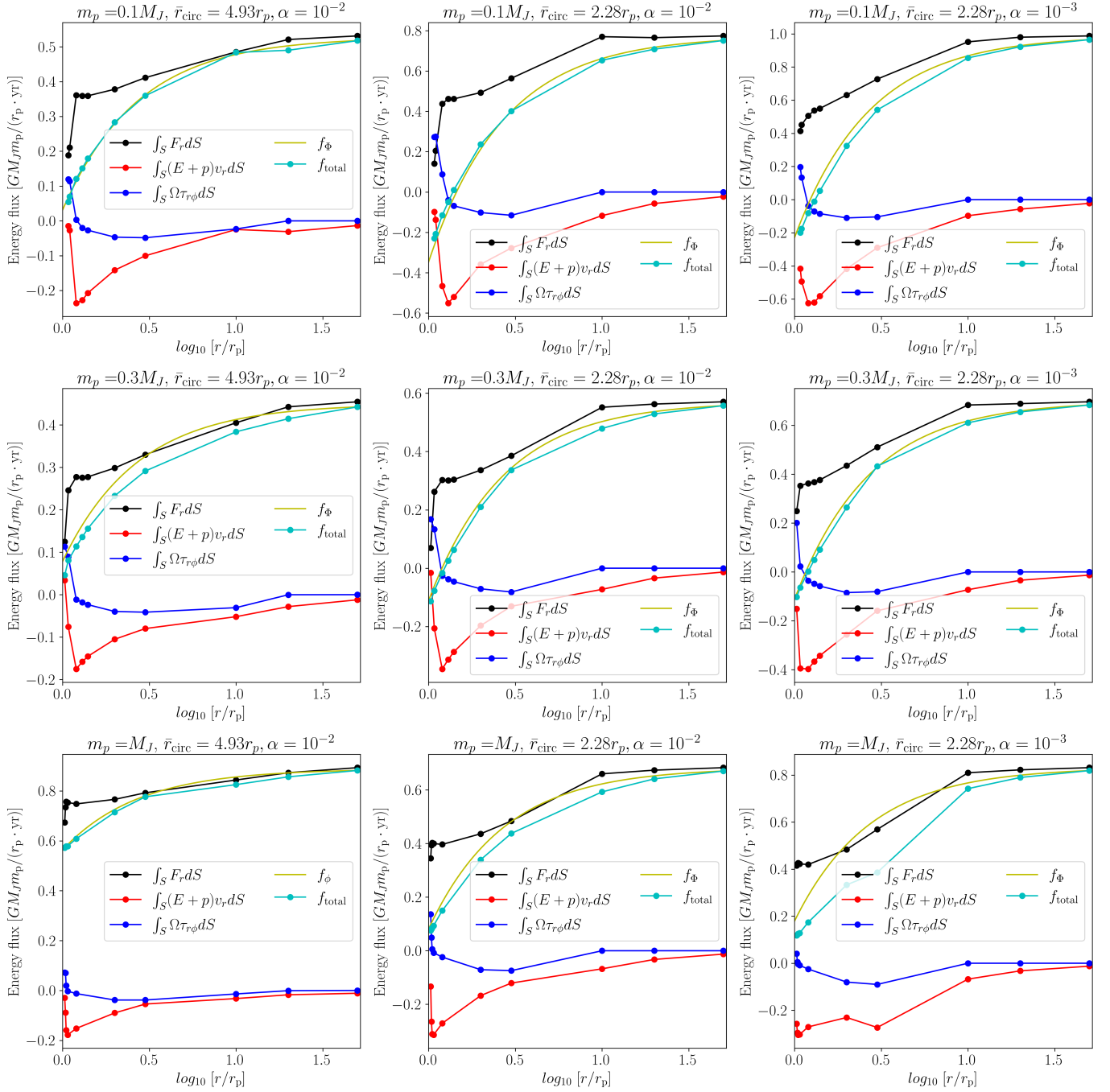}
    \caption{Energy fluxes of all simulations. First row: models 1, 2, and 3. Second row: models 4, 5, and 6. Third row: models 7, 8, and 9.}
    \label{fig:energyfluxensemble}
\end{figure*}

Figure \ref{fig:energyfluxensemble} shows the energy fluxes of all 9 simulations. Most of them exhibit good energy conservation within the computational domain, with close-to-zero $f_{\rm total}$---i.e., a small net energy flux---at the inner boundaries, and reveal the physical energy conversion between different energy forms. The simulations with $\mpl=\mj$ are the most expensive ones; model 9 may not have reached the quasi-steady state. Its mass accretion rates at $r=\{1.2,1.5,1.8\}\,\rpl$ are still evolving, and there is some difference between $f_{\Phi}$ and $f_{\rm{total}}$. In addition, model 7 seems to have a large energy flux into the computational domain, which may lead to a higher temperature in the simulation.

Finally, despite efforts to optimize energy conservation, exact conservation remains unattainable under such complex physical conditions owing to the explicit source terms for gravitational and inertial forces (Equation \ref{eqn:sphereE}). Additionally, the implicit radiation transport solver introduces minor conservation errors. Further improving the model's numerical accuracy remains a subject for future work.

\end{CJK*}
\end{document}